\documentclass[a4paper]{elsarticle}
\usepackage{graphicx}
\usepackage[english]{babel}
\usepackage{here}
\usepackage{subfloat}
\usepackage{fontenc}
\usepackage[utf8]{inputenc}
\usepackage[colorlinks=false]{hyperref}
\usepackage[printonlyused]{acronym}
\usepackage{bm}
\usepackage{bbold}
\usepackage{amsmath}
\usepackage{slashed}
\usepackage{amssymb}
\usepackage{textcomp}
\usepackage[usenames]{color}
\usepackage{enumerate}
\usepackage{dcolumn}
\usepackage{longtable}
\usepackage[usenames]{color}
\newcommand{\bfigh}{\begin{figure}[H]}					

\newcommand{\bfig}{\begin{figure}}		

\newcommand{\efig}{\end{figure}}

\newcommand{\ra}{\rangle}

\newcommand{\beqn}{\begin{eqnarray*}}
\newcommand{\eeqn}{\end{eqnarray*}}
\newcommand{\beq}{\begin{eqnarray}}
\newcommand{\eeq}{\end{eqnarray}}
\newcommand{\bea}{\begin{eqnarray}}
\newcommand{\eea}{\end{eqnarray}}
\newcommand{\be}{\begin{equation}}
\newcommand{\ee}{\end{equation}}

\newcommand{\bmatr}{\begin{pmatrix}}
\newcommand{\ematr}{\end{pmatrix}}
\definecolor{Red}{rgb}{1,0,0}

\newcommand{\Psib}{\overline{\Psi}}

\newcommand{\Lcal}{{\cal L}}
\newcommand{\p}{\partial}
\newcommand{\se}{\Sigma}

\newsavebox{\ZitName}

\begin{document}
\title{SU(3) Approach to Hypernuclear Interactions and Spectroscopy}

\author{Horst \textsc{Lenske}$^{1,2}$, Madhumita \textsc{Dhar}$^{1}$, Theodoros \textsc{Gaitanos}$^{1,3}$\\
$^{1}$Institut f\"ur Theoretische Physik, JLU Giessen, Heinrich-Buff-Ring 16, D-35392, Giessen, Germany \\
$^2$GSI Darmstadt, D-64291 Darmstadt, Germany\\
$^{3}$Aristotle University of Thessaloniki, Thessaloniki, Greece}



%
%
\date{\today}
%
\begin{abstract}

The $SU(3)$ meson exchange approach to interactions within the baryon octet and nuclear density functional theory are used to derive an \emph{ab initio} description of hypernuclear interactions. The density dependence of interactions is recast into a DFT with density dependent interaction vertices. The field-theoretical structure is retained by expressing the vertices as functionals of the matter field operators. Applications to infinite hypermatter and neutron star matter are discussed. A new approach is presented allowing to determine in-medium coupling constants out of the $NN$-vertex functionals, obtained e.g. by DBHF theory, for the full baryon octet by exploiting $SU(3)$ relations.
\end{abstract}
\maketitle

\section{Introduction}\label{sec:intro}

Hypernuclear physics has obtained new momentum by a series of new observation of an exotic system like $^6_\Lambda H$ \cite{Agnello:2012jw}, an unexpected and yet unexplained short-life time of the hypertriton \cite{Rappold:2014jqa}, and strong indications for a $nn\Lambda$ bound state \cite{HypHI} which would be the first and hitherto only charge-neutral system bound by strong interactions. Recent results on light hypernuclei and their antimatter counterparts \cite{Shah:2015oha,She:2015bta} at RHIC and the LHC, respectively, seem to confirm the surprising life-time reduction and, moreover, point to a not yet understood reaction mechanism. The "hyperonization puzzle" heavily discussed for neutrons stars \cite{Bednarek:2011gd} is another aspect in the revived strong interest in in-medium strangeness physics.

Hyperons and hypernuclei are short-lived objects and as such do not exist as stable particles. Their production always requires special efforts, as reviewed e.g. in \cite{Hash:2006,Tamura:2013,Lenske:2015}. High energy hadronic reactions, providing a broad spectrum of final fragments, are a very suitable tool for investigations of strangeness in the nuclear medium. Our recent studies \cite{Shyam:2004,Shyam:2006,Bender:2009,Gaitanos:2008uc,Gaitanos:2009,Gaitanos:2011,Larionov:2012,Gaitanos:2012,Gaitanos:2014} of the production mechanism in the Giessen resonance model are emphasizing the role played by baryon resonances as the most important source for strangeness and hypernuclear production, at least at energies below $T_{Lab}\sim 10$~GeV. A report on our latest results on strangeness production will be found elsewhere in this volume \cite{Gaitanos:2016}. A typical scenario is that a target nucleon is excited into a nucleon resonance $N^*$, being located above the strangeness production threshold and decaying into a hyperon $Y$ and an antikaon $\bar K$. This resonance mechanism is an important scenario in hadron-induced hyperon production as well as in strangeness production in heavy ion collisions. The dense and hot environment in a heavy ion collision at energies of a few GeV per nucleon allows additional production paths by secondary rescattering processes of hadrons already produced in earlier stages of the reaction. Those results lead to the conclusion that hadronic degrees of freedom are prevailing in nuclear matter, even at densities a few times the nuclear saturation density $\rho_{sat}=0.16fm^{-3}$ and moderate temperatures. Hence, a description of hypermatter and hypernuclei by baryons and mesons is well justified.

Hypernuclear production processes are intimately connected to the interaction of hyperons with the background medium by setting the conditions for the formation or non-existence of hypernuclear bound states \cite{Lenske:2015}. While a larger number of $\Lambda$-hypernuclei are known, see e.g. \cite{Hash:2006}, no safe signal for a particle-stable $\Sigma$ or a $S=-2$ Cascade hypernucleus has been recorded. However, few examples of double-$\Lambda$ hypernuclei have been reported. Thus, a unified approach covering the interactions and dynamics within the lowest baryon flavour octet is a desirable and worthwhile attempt. In light nuclei Fadeev-methods allow an \emph{ab initio} description by using free space baryon-baryon ($BB$) potentials directly as practiced e.g. in \cite{Hiyama:2013owa}. For heavier nuclei, density functional theory ($DFT$) is the only available approach being applicable over wide ranges of nuclear masses.  Already some time ago we have made first steps in such a direction \cite{Keil:2002} within the Giessen Density Dependent Hadron Field ($DDRH$) theory. The DDRH approach incorporates Dirac-Brueckner Hartree-Fock ($DBHF$) theory into covariant density functional theory ($DFT$) \cite{Lenske:1995,Fuchs:1995,Hofmann:2001,Keil:2003,Lenske:2004,Fedo:2014}. Since then, the approach is being used widely on a purely phenomenological level as e.g. in  \cite{Typel:1999yq,Lala:2005,Liang:2014dma,Niksic:2014qsa}. In the non-relativistic sector comparable attempts are being made, ranging from Brueckner theory for hypermatter \cite{Schulze:1998jf} to phenomenological density functional theory extending the Skyrme-approach to hypernuclei \cite{Lanskoy:1997xq,Schulze:2014oia}.

Nucleons and hyperons together are forming the ground state baryon $SU(3)$ flavour octet. SU(3) symmetry, however is broken as is obvious from the spread of about $400$~MeV among the octet baryon masses, corresponding to a window of $\pm 20$\% around the mean mass of the baryon octet. Of course, broken SU(3) symmetry is taken onto account theoretically, at least on the level of masses and the corresponding production thresholds. In this work, we study formal aspects of octet baryon-meson interactions in section \ref{sec:OctetInter}. New theoretical results for $BB'$-scattering in free space are presented in section \ref{sec:FreeSpace}. The evolution of interactions in nuclear matter is studied in section \ref{sec:BB_Infinite}.  An update of Giessen DDRH density functional approach is discussed in section \ref{sec:HyperDFT} and recent results for infinite hypermatter and $\Lambda$~-hypernuclei are presented. The mapping of in-medium interactions to an energy density functional with density dependent interaction vertices is investigated with special focus on the diagrammatic structure of in-medium coupling constants. Exploiting the SU(3)-relations among octet coupling constants, we present in section \ref{sec:Octet_Interactions} a new approach predicting in-medium vertices for the whole baryon octet. The work is summarized in section \ref{sec:SumLook} and an outlook to future research is given.

\section{Interactions in the baryon octet}\label{sec:OctetInter}

The statement that the quantitative understanding of hypernuclei requires an accurate knowledge of hyperon-hyperon $(YY)$ and hyperon-nucleon $(YN)$ interactions sounds obvious but, in fact, is far from being trivial. The solution of this basic task is still far from being under control to a satisfactory level at the degree of accuracy required for a hypernuclear theory of predictive power. The progress made in the last decade or so for $S=-1$ systems is only part of the full story since this involves mainly single-$\Lambda$ hypernuclei, supplementing the few data points from old $p\Lambda$ and $p\Sigma$ experiments. The latter are essential input for the approaches developed over the years by several groups. While the Nijmegen \cite{Nijmegen,Yama:2014} and the Juelich group \cite{Juelich}, respectively, are using a baryon-meson approach, the Kyoto-Niigata group \cite{Niigata} favors a quark-meson picture, finally ending also in meson-exchange interactions. A QCD-inspired path is followed by chiral effective field theory, reviewed e.g. in Ref. \cite{Epelbaum:2008}. However, none of the existing parameter sets is in any sense constrained with respect to interactions in the $|S|\geq 2$ channels. As a consequence, the implementation of $SU(3)$ flavour symmetry is incomplete. In view of this situation, we found it necessary to reconsider interactions in the lowest $SU(3)$ baryon octet with special emphasis on in-medium interactions. As a long-term program the approach will allow to test interactions more directly under the dynamical conditions of strangeness and hypernuclear production reactions \cite{Lenske:2015,Gaitanos:2014,Gaitanos:2016} with leptonic, hadronic, and nuclear probes, up to heavy ion reactions at relativistic energies. Hyperon and hypernuclear data from reactions are providing a wealth of information which has not yet been taken advantage of in full breadth.

\begin{figure}[tbh]
\begin{center}
\includegraphics[width=8cm]{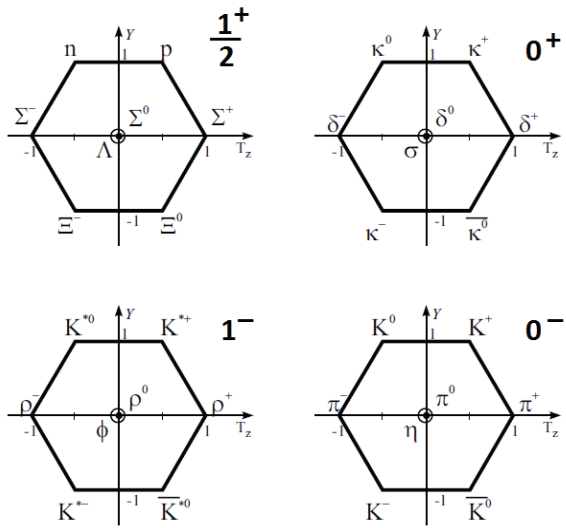}
\caption{The $SU(3)$ multiplets considered in this work: the $\frac{1}{2}^+$ baryon octet (upper left), the pseudo-scalar $0^+$ (upper right), the vector $1^-$ (lower left), and the pseudo-scalar $0^-$ (lower right). The octets have to be combined with the corresponding meson-singlets, which are not displayed, thus giving rise to meson nonets.}
\label{fig:Octets}
\end{center}
\end{figure}

In Fig. \ref{fig:Octets} the $SU(3)$ multiplets are shown which are considered in our approach. The lowest $\frac{1}{2}^+$ baryon octet, representing the baryonic ground state multiplet, is taken into account together with three meson multiplets, namely the pseudo-scalar ($P$), the vector ($V$), and the scalar ($S$) octets. In addition, we include also the corresponding meson singlet states, represented physically by the $\eta', \phi, \sigma'$ states but not shown in Fig. \ref{fig:Octets}. As discussed below, the observed isoscalar-octet and -singlet mesons are in fact mixtures of bare SU(3) octet-states $\eta_8,\omega_8,\sigma_8$ and singlet states $\eta_1,\phi_1,\sigma_1$. The mesons $\alpha$ of each multiplet couple to the baryons by vertices of a typical, generic structure,
\bea
\Gamma_\alpha(0^-)&:&\quad  g_{BB'\alpha}\bar \Psi_{B'}\gamma_5\gamma_\mu\Psi_B\partial^\mu\Phi\\
\Gamma_\alpha(1^-)&:&\quad  g_{BB'\alpha}\bar \Psi_{B'}\gamma_\mu\Psi_B\partial^\mu\Phi
+\frac{f_{BB'\alpha}}{M_B+M_{B'}}\bar{\Psi}_{B'}\sigma_{\mu\nu}\Psi_BF^{\mu\nu}_\alpha\\
\Gamma_\alpha(0^+)&:&\quad  g_{BB'\alpha}\bar \Psi_{B'}\Psi_B\Phi\\
\eea
given by Lorentz-covariant bilinears of baryon field operators $\Psi_B$ and the Dirac-conjugated field operators $\bar{\Psi}_B=\gamma_0\Psi^\dag_B$ and Dirac $\gamma$~-matrices, to which meson fields ($V,S$) or their derivatives ($P$) are attached, such that in total a Lorentz-scalar is obtained. Vector mesons and baryons may also couple via the relativistic tensor operator $\sigma_{\mu\nu}=\frac{i}{2}[\gamma_\mu,\gamma_\nu]$, given by commutators of $\gamma$ matrices, and separate tensor coupling constant $f_{BB'\alpha}$. Here, we neglect the tensor coupling, except for a discussion of spin-orbit effects in section \ref{sec:HyperDFT}. The Bjorken-convention \cite{Bjorken:1964} is used for spinors and $\gamma$-matrices.

\begin{table}
\centering
\begin{tabular}{|c|l|c|c |}
\hline
Channel & Meson & Mass [MeV]  & Cut-off [MeV/c]\\\hline
$0^{-}$ & $\pi$ &138.03 & 1300\\ \hline
$0^{-}$ & $\eta $ & 547.86 & 1300\\ \hline
$0^{-}$ &$ K^{0,+} $  & 497.64 & 1300 \\ \hline
$0^{+}$ & $\sigma $ & 760.0& 1850 \\ \hline
$0^{+}$ & $\delta $ & 983.0 & 2000\\ \hline
$0^{+}$ & $ \kappa $ & 880.0 & 2000\\ \hline
$1^{-}$ & $\omega $ & 782.65& 1700\\ \hline
$1^{-}$ & $\rho $ & 775.26 & 1700\\ \hline
$1^{-} $& $K^{*} $ & 891.66 & 1700\\
 \hline
\end{tabular}\caption{Meson masses and cut-off momenta treated as fixed valued-input in the calculations. The scalar meson mass is chosen here in the higher energy tail of the $f_0(500)$ spectral distribution.}
	\label{tab:OBE_Param}
\end{table}

While the pseudo-scalar and the vector mesons are well identified as physical particles or poles in the complex plane, the situation is less clear for the scalar nonet. We shall come back to this question in a later section. Here, we mention that we identify $\sigma=f_0(500)$, $\delta=a_0(980)$, and $\kappa=K^*_0(800)$ as found in the compilations of the Particle Data Group \cite{PDG:2012}. In the numerical calculations, we use an isoscalar-scalar mass $m_\epsilon = 760$~MeV, roughly corresponding to the $\epsilon$~-meson of the Nijmegen-group, still being located well within the broad spectral distribution of $f_0(500)$. The so-called $\kappa$-meson has been observed rather only recently as resonance-like structures in charmonium decay spectroscopy. A two-bump structure with maxima at about $640$~MeV and $800$~MeV has been observed. Since their recent compilation PDG recommends a mean mass $m_\kappa=682\pm 29$~MeV \cite{PDG:2012}. The meson parameters are summarized in Tab. \ref{tab:OBE_Param}, the octet baryons are listed in Tab. \ref{tab:Baryons}.

\begin{table}
\centering
\begin{tabular}{|l|r|l|r|}
\hline
Baryon       & Mass [MeV]& Baryon       & Mass [MeV]\\ \hline
n            & 938.56563 & $\Sigma^{-}$ & 1197.436  \\ \hline
p            & 1877.27231& $\Sigma^{0}$ & 1192.55   \\ \hline
$\Lambda$    & 1115.684  & $\Xi^{-}$    & 1321.7    \\ \hline
$\Sigma^{+}$ & 1189.37   & $\Xi^{0}$    & 1314.86   \\ 
 \hline
\end{tabular}
\caption{The octet baryon masses.}
	\label{tab:Baryons}
\end{table}

We introduce the flavour spinor $\Psi_F$
\beq
\Psi_F = \left( \Psi_N, \Psi_\Lambda, \Psi_\Sigma, \Psi_\Xi \right)^{\text{T}}
\label{eq:FlavourSp}
\eeq
being composed of the isospin multiplets
\beq
\Psi_N = \left( \begin{array}{c} \psi_p \\ \psi_n \end{array} \right),\quad
\Psi_\Lambda = \psi_\Lambda,\quad
\Psi_\Sigma = \left( \begin{array}{c}
\psi_{\Sigma^+} \\ \psi_{\Sigma^0} \\ \psi_{\Sigma^-} \end{array} \right),\quad
\Psi_\Xi = \left( \begin{array}{c}
\psi_{\Xi^0} \\ \psi_{\Xi^-} \end{array} \right)
\eeq
where $\psi_i$ are Dirac spinors. The full Lagrangian is structured in an
isospin symmetric way, bing in total an isospin scalar. We are dealing with the Lagrangian
\be\label{Lagrangian}
\Lcal = \Lcal_{B} + \Lcal_{M} + \Lcal_{int}
\ee
accounting for the free motion of baryons with mass matrix $\hat{M}$,
\be
\Lcal_{B} = \Psib_{F} \left[ i\gamma_\mu\p^\mu
                              - \hat{M} \right] \Psi_{F} \quad ,
\ee
and the Lagrangian density of massive mesons,
\be
\Lcal_{M} =\frac{1}{2} \sum_{i = P,S,V}
\left(\p_\mu\Phi_i\p^\mu\Phi_i - m_{\Phi_i}^2\Phi_i^2\right)
           - \frac{1}{2} \sum_{\kappa = V}
             \left( \frac{1}{2} F^{(\kappa)^2} - m_\kappa^2 V^2_{\kappa}
\right)
\ee
where $P,S,V$ denote summations over the lowest nonet pseudo-scalar, scalar, and vector mesons. Of special interest for nuclear matter and nuclear structure research are the mean-field producing meson fields, involving isoscalar-scalar mesons $\sigma$, $\sigma_s$, the isovector-scalar $\delta=a_0(980)$ meson, and their isoscalar-vector counterparts $\omega,\phi$ and the isovector-vector $\rho$ meson, respectively. For finite nuclei, the photon field $V^\mu_\gamma$ is included. The field strength tensor of the vector meson fields $V^\mu_\kappa$, $\kappa\in\{\omega,\rho,K^*,\phi,\gamma\}$ is given by
\beq
F^{(\kappa)}_{\mu\nu} = \p_\mu V_\nu^{(\kappa)} - \p_\nu V_\mu^{(\kappa)} \quad .
\label{eq:Fmunu}
\eeq
Baryon-meson interactions, contained in $\Lcal_{int}$, are specified below.

$SU(3)$ octet physics is based on treating the eight baryons on equal footing. Although $SU(3)$ flavour symmetry is broken by about 20\% on the mass scale, it is still meaningful to exploit the relations among coupling constants, thus defining a guideline and reducing the number of free parameters considerably. The eight $J^P={\textstyle\frac{1}{2}}^+$ baryons are collected into a traceless matrix $B$, which is given by a superposition of the eight Gell-Mann matrices $\lambda_i$, leading to the familiar form
\begin{equation}
  B =\sum_{i=1\cdots 8}{\lambda_i B_i}= \left( \begin{array}{ccc}
      {\displaystyle\frac{\Sigma^{0}}{\sqrt{2}}+\frac{\Lambda}{\sqrt{6}}}
               &  \Sigma^{+}  &  p  \\[2mm]
      \Sigma^{-} & {\displaystyle-\frac{\Sigma^{0}}{\sqrt{2}}
                   +\frac{\Lambda}{\sqrt{6}}}  &  n \\[2mm]
      -\Xi^{-} & \Xi^{0} &  {\displaystyle-\frac{2\Lambda}{\sqrt{6}}}
             \end{array} \right),
\end{equation}
and which is invariant under $SU(3)$ transformations. The pseudo-scalar($P$), vector ($V$), and, last but not least, the scalar ($S$) meson nonets are constructed correspondingly. Taking the $J^P=0^-$ pseudo-scalar mesons as an example we obtain the octet matrix $P_{8}$
\begin{equation}
   P_{8} = \left( \begin{array}{ccc}
      {\displaystyle\frac{\pi^{0}}{\sqrt{2}}+\frac{\eta_{8}}{\sqrt{6}}}
             & \pi^{+}  &  K^{+}  \\[2mm]
      \pi^{-} & {\displaystyle-\frac{\pi^{0}}{\sqrt{2}}
         +\frac{\eta_{8}}{\sqrt{6}}}  &   K^{0} \\[2mm]
      K^{-}  &  \overline{{K}^{0}}
             &  {\displaystyle-\frac{2\eta_{8}}{\sqrt{6}}}
             \end{array} \right).
\end{equation}
which, for the full nonet, has to be completed by the singlet matrix $P_1$, given by the unit matrix multiplied by $\eta_0/\sqrt{3}$. We define the $SU(3)$-invariant combinations
\begin{eqnarray}
  \left[\overline{B}BP_8\right]_{D}&=& {\rm Tr}\left(\left\{\overline{B},B\right\}P_{8}\right) \quad , \quad
  \left[\overline{B}BP_8\right]_{F} = {\rm Tr}\left(\left[\overline{B},B\right]P_{8}\right)
\quad , \nonumber \\
\left[\overline{B}BP_1\right]_{S} &=& {\rm Tr}(\overline{B}B){\rm Tr}(P_{1})/\sqrt{3}
\end{eqnarray}
where $F$ and $D$  correspond to anti-symmetric and symmetric combinations of field operators, respectively, as indicated by the commutators $[X,Y]$ and anti-commutators, $\{X,Y\}$, respectively. The singlet interaction term is denoted by $S$, normalized such that the meson trace is unity. With these relations we obtain the interaction Lagrangian
\begin{equation}
   {\cal L}_{I} =  - \left\{
     g_D\left[\overline{B}BP_8\right]_{D} + g_F\left[\overline{B}BP_8\right]_{F} \right\}\, - \,
     g_S\left[\overline{B}BP_1\right]_{S},             \label{eq:LIsu3}
\end{equation}
with the generic $SU(3)$ coupling constants $\{g_{D},g_F,g_S\}_P$. For later use, we introduce the isospin doublets
\begin{equation}
  N=\left(\begin{array}{c} p \\ n \end{array} \right), \ \ \
  \Xi=\left(\begin{array}{c} \Xi^{0} \\ \Xi^{-} \end{array} \right), \ \ \
  K=\left(\begin{array}{c} K^{+} \\ K^{0} \end{array} \right),
  \ \ \   K_{c}=\left(\begin{array}{c} \overline{K^{0}} \\
               -K^{-} \end{array} \right) \quad .        \label{eq:doublets}
\end{equation}
The $\Sigma$ hyperon and the pion isovector-triplets are expressed in the basis defined by the spherical unit vectors $\bm{e}_{\pm,0}$ which leads to
\begin{equation}
  \bm{\Sigma}\!\cdot\!\bm{\pi} = \Sigma^{+}\pi^{-}+\Sigma^{0}\pi^{0}+\Sigma^{-}\pi^{+} \quad ,
\end{equation}
also serving to fix phases. We define the pseudo-vector Dirac-vertex operator $\Gamma=\gamma_{5}\gamma_{\mu}\partial^{\mu}$. By evaluating the $F$- and $D$-type couplings, Eq.~(\ref{eq:LIsu3}), we obtain the pseudo-scalar octet-meson interaction Lagrangian in an obvious, condensed short-hand notation, following the pioneering work of de Swart \cite{deSwart:1963}
\begin{eqnarray}
  m_{\pi}{\cal L}_{8} &\sim& \nonumber \\
  &-&g_{NN\!\pi}(\overline{N}\Gamma\bm{\tau}N)\!\cdot\!\bm{\pi}
  +ig_{\Sigma\Sigma\pi}(\overline{\bm{\Sigma}}\!\times\!\Gamma\bm{\Sigma})
      \!\cdot\!\bm{\pi}\nonumber \\
  &-&g_{\Lambda\Sigma\pi}(\overline{\Lambda}\Gamma\bm{\Sigma}+
      \overline{\bm{\Sigma}}\Gamma\Lambda)\!\cdot\!\bm{\pi}
  -g_{\Xi\Xi\pi}(\overline{\Xi}\Gamma\bm{\tau}\Xi)\!\cdot\!\bm{\pi}
            \nonumber\\
 &-&g_{\Lambda N\!K}\left[(\overline{N}\Gamma K)\Lambda
         +\overline{\Lambda}\Gamma(\overline{K}N)\right] \nonumber \\
 &-&g_{\Xi\Lambda K}\left[(\overline{\Xi}\Gamma K_{c})\Lambda
         +\overline{\Lambda}\Gamma (\overline{K_{c}}\Xi)\right] \nonumber\\
 &-&g_{\Sigma N\!K}\left[\overline{\bm{\Sigma}}\!\cdot\Gamma\!
         (\overline{K}\bm{\tau}N)+(\overline{N}\Gamma\bm{\tau}K)
         \!\cdot\!\bm{\Sigma}\right] \nonumber \\
 &-&g_{\Xi\Sigma K}\left[\overline{\bm{\Sigma}}\!\cdot\!
       \Gamma (\overline{K_{c}}\bm{\tau}\Xi)
     +(\overline{\Xi}\Gamma\bm{\tau}K_{c})\!\cdot\!\bm{\Sigma}\right]
                                         \nonumber\\
 &-&g_{N\!N\eta_{8}}(\overline{N}\Gamma N)\eta_{8}
   -g_{\Lambda\Lambda\eta_{8}}(\overline{\Lambda}\Gamma\Lambda)\eta_{8} \nonumber \\
 &-&g_{\Sigma\Sigma\eta_{8}}(\overline{\bm{\Sigma}}\!\cdot\!\Gamma
       \bm{\Sigma})\eta_{8}
   -g_{\Xi\Xi\eta_{8}}(\overline{\Xi}\Gamma\Xi)\eta_{8}.    \label{eq:Lbar8}
\end{eqnarray}
where the charged-pion mass serves as the mass-scale compensating the effect of the derivative coupling. The - in total 16 - pseudo-scalar $BB'$-meson vertices are completely fixed by the three nonet coupling constants $(f_8,f_1,\alpha)$ or, likewise, by $(g_{D},g_{F},g_S)_P$.

Corresponding relations exist also for the vector and the scalar nonets, $V^\mu$ and $S$, respectively, with their own sets of respective coupling constants $(g_{D},g_{F},g_{S})_{s,v}$. \footnote{We neglect the rank-2 tensor $BB'$-vector meson coupling} The vector case is obtained by the mapping $\{K,\pi,\eta_8,\eta_1\}\to \{K^*,\rho,\omega_8,\phi_1\}$ and the scalar couplings are obtained by replacing $\{K,\pi,\eta_8,\eta_1\}\to \{\kappa,\delta,\sigma_8,\sigma_1\}$. The coupling constants of the three nonets, ($m=P,V,S$), obey a common structure. Denoting the isoscalar octet meson by $f\in \{\eta_8,\omega_8,\sigma_8 \}$, the isovector octet meson by $a\in \{\pi,\rho,\delta \}$, and the isospin doublet meson by $K\in \{K^{0,+},K^{*0,+},\kappa=K^{*,0,+}_0 \}$, the interaction strengths are determined by the relations
\begin{equation}
  \begin{array}{lll}
   g_{N\!N a}               = \sqrt{2}(g_{D}+g_{F}),                               \ \ \  &
   g_{\Lambda N\!K}          =-\sqrt{\frac{2}{3}}(g_{D}+2g_{F}),\ \ \  &
   g_{N\!Nf}          = \frac{1}{\sqrt{6}}(3g_{F}-g_{D}),
      \\
   g_{\Sigma\Sigma a}       = \sqrt{2}g_{F},                        \ \ \  &
   g_{\Xi\Lambda K}          = \sqrt{\frac{2}{3}}(3g_{F}-g_{D}),\ \ \  &
   g_{\Lambda\Lambda f} =-\sqrt{\frac{2}{3}}g_{D},
      \\
   g_{\Lambda\Sigma a}      = \sqrt{\frac{2}{3}}g_{D}, \ \ \  &
   g_{\Sigma N\!K}           = \sqrt{2}(g_{D}-g_{F}),                    \ \ \  &
   g_{\Sigma\Sigma f}  = \sqrt{\frac{2}{3}}g_{D},
      \\
   g_{\Xi\Xi a}             =-\sqrt{2}(g_{D}-g_{F}),                    \ \ \  &
   g_{\Xi\Sigma K}           =-2(g_{D}+g_{F}),                               \ \ \  &
   g_{\Xi\Xi f}        =-\frac{1}{\sqrt{6}}(3g_{F}+g_{D}).
  \end{array}
                        \label{eq:goct}
\end{equation}
where $\{g_{D},g_F,g_S\}$ denotes either set of pseudo-scalar, vector, or scalar couplings. The interactions due to exchange of the isoscalar-singlet meson $f'\in \{\eta_1,\phi_1,\sigma_1 \}$  are treated accordingly with the result
\begin{equation}
   g_{N\!N f'}=g_{\Lambda\Lambda f'} =g_{\Sigma\Sigma f'}=g_{\Xi\Xi f'} = g_{S}.  \label{eq:gsin}
\end{equation}

\subsection{Dynamical $SU(3)$ symmetry breaking}\label{ssec:SymBreak}
From the above relations, the advantage of referring to $SU(3)$ symmetry is obvious: For each type of meson (pseudo-scalar, vector, scalar) only four independent parameters are required to characterize their interaction strengths with all possible baryons. These are the singlet
coupling constant $g_S$, the octet coupling constants $g_D,g_F$, and eventually the three mixing angles $\theta_{P,V,S}$, one for each meson multiplet, which relate the physical, dressed isoscalar mesons to their bare octet and singlet counterparts. Likewise, a different set of parameters is used frequently to describe the $(P,V,S)$ multiplets, given by $f_8=g_D+g_F$, the ratios $\alpha_8=g_F/(g_F+g_D)$, and the singlet coupling constant $f_1=g_S$. It should also be noted that representations are in use extracting factors $\sqrt{2}$ and $\frac{1}{\sqrt{3}}$ from the ($D,F$)-type and the $S$-type Lagrangians, respectively.

$SU(3)$ symmetry, however, is broken at several levels and $SU(3)$ relations will not be satisfied exactly. An obvious one is the non-degeneracy of the physical baryon and meson masses within the multiplets. As discussed below, this splitting leads to additional complex structure in the set of coupled equations for the scattering amplitudes because the various baryon-baryon ($BB'$) channels open at different threshold energies $\sqrt{s_{BB'}}=m_B+m_{B'}$. At energies $s<s_{BB'}$ a given $BB'$-channel does not contain asymptotic flux but contributes as a virtual state. Thus, $N\Lambda$ scattering, for example, will be modified at any energy by admixtures of $N\Sigma$ channels.

In addition, $SU(3)$ symmetry is broken explicitly by the fact that $\Lambda$ and $\Sigma^0$ have the same quark content but coupled to different total isospin. There is an appreciable electro-weak mixing between the ideal isospin-pure $\Lambda$ and $\Sigma^0$ states~\cite{Dal64}. Exact $SU(3)$ symmetry of strong interactions predicts $f_{\Lambda\Lambda\pi^0}=0$, but $\Lambda$-$\Sigma^0$ mixing results in a weak, but non-zero $\Lambda\Lambda\pi$ coupling constant for the physical $\Lambda$-hyperon, as derived by Dalitz and von Hippel~\cite{Dal64} already in the early days of strangeness physics,
\begin{equation}
  g_{\Lambda\Lambda\pi}=-2\frac{\langle\Sigma^0|\delta M|\Lambda\rangle}
             {M_{\Sigma^0}-M_{\Lambda}}\,f_{\Lambda\Sigma\pi},
\end{equation}
where the $\Sigma\Lambda$ element of the mass matrix is given by
\begin{equation}
   \langle\Sigma^0|\delta M|\Lambda\rangle=
       \left[M_{\Sigma^0}-M_{\Sigma^+}+M_p-M_n\right]/\sqrt{3}.
\end{equation}
Substituting the physical baryon masses, we find
$ g_{\Lambda\Lambda\pi}=c_b\,g_{\Lambda\Sigma\pi}$ where, up to four digits, $c_b=-0.0283$ is the symmetry breaking coefficient. From the nucleon-nucleon-pion part of the interaction Lagrangian, Eq.~(\ref{eq:Lbar8}), we find the isospin matrix element
\begin{equation}
    (\overline{N}\bm{\tau}N)\!\cdot\!\bm{\pi} =
    \overline{p}p\pi^0-\overline{n}n\pi^0+\sqrt{2}\,\overline{p}n\pi^+
                      +\sqrt{2}\,\overline{n}p\pi^-,
\end{equation}
and the neutral pion is seen to couple with opposite sign to neutrons and protons. This implies that the non-zero $f_{\Lambda\Lambda\pi^0}$ coupling produces considerable deviations from charge symmetry in $\Lambda p$ and $\Lambda n$ interactions. Obviously, $\Lambda$-$\Sigma^0$ mixing also gives rise to non-zero $\Lambda\Lambda$ coupling constants for any (neutral) isovector meson, in particular for the $\rho$ vector meson and the $\delta/a_0(980)$, the isovector member of the $0^+$ meson octet, respectively but the latter couplings give rise to much smaller effects.

As seen from Eq.~(\ref{eq:goct}), $\Lambda$ and $\Sigma$ hyperons are coupled by strong interactions leading to a strangeness- and isospin/charge-conserving dynamical mixing of $N\Lambda$ and $N\Sigma$ channels. Again, corresponding effects are present also for vector-isovector and scalar-isovector mesons. As a result, channels of the same total strangeness $S$ and fixed total charge $Q$ are coupled already on the level of the elementary vertices in free space. In asymmetric nuclear matter, this effect is enhanced through the isovector mean-field produced by the $\rho_0$ and the $\delta_0$ meson fields \cite{Lenske:2015,Muller:1998pr}. This leads to a density dependent $\Lambda\Sigma^0$ mixing being proportional to the isovector mean-field $U_1\sim \rho_1$ and determined by the isovector baryon density $\rho_1=\rho_n-\rho_p$, to be supplemented by possible contributions from charged hyperons. As a result, $\Lambda$ and $\Sigma^0$ become quasi-particles with mass eigenstates different from the flavour eigenstates. The total isospin is of course conserved by virtue of the background medium.

\begin{table}[]
\begin{center}
\begin{tabular}{|c|c|c|c|c|c|}
	\hline
   & \textbf{Q=-2} & \textbf{Q=-1}  & \textbf{Q=0}& \textbf{Q=1} & \textbf{Q=2} \\ \hline
   \textbf{ S=0}   & & &nn  & np & pp\\\hline
\textbf{S= -1} & & $\Sigma^{-}n$  & \begin{tabular}{@{}c@{}c@{}}$\Lambda n$ \\ $\Sigma^{0} n$\\ $\Sigma^{ -}p $ \\ \end{tabular} &  \begin{tabular}{@{}c@{}c@{}}$\Lambda p$ \\ $\Sigma^{+ }n$\\ $\Sigma^0 p$ \\\end{tabular}  & $\Sigma^{+}p$\\  \hline
    \textbf{S=-2}  & $\Sigma^{-}\Sigma^{-} $ &  \begin{tabular}{@{}c@{}c@{}} $\Xi^{-}n$ \\ $\Sigma^{-}\Lambda$ \\ $\Sigma^{-} \Sigma^{0}$ \\ \end{tabular}  & \begin{tabular}{@{}c@{}c@{}c@{}c@{}c@{}}$\Lambda \Lambda$ \\$\Xi^{0}$n \\$\Xi^{-}$p\\ $\Sigma^{+}\Lambda$\\$ \Sigma^{+} \Sigma^{0} $\\ \end{tabular} & \begin{tabular}{@{}c@{}c@{}} $\Xi^{0}$p \\ $\Sigma^{+}\Lambda$\\ $ \Sigma^{+} \Sigma^{0} $ \\ \end{tabular} &  $\Sigma^{+} \Sigma^{+}$\\  \hline
   \textbf{ S= -3} &  $\Xi^{-} \Sigma^{-}$&  \begin{tabular}{@{}c@{}c@{}} $\Xi^{-} \Lambda$ \\  $\Xi^{0} \Sigma^{-}$\\   $\Xi^{-}\Sigma^{0} $ \\ \end{tabular}  & \begin{tabular}{@{}c@{}c@{}} $\Xi^{0} \Lambda$\\ $ \Xi^{0}  \Sigma^{0}$\\   $\Xi^{-}\Sigma^{+} $\\ \end{tabular}  &   $\Xi^{0}\Sigma^{+}$  & \\  \hline
    \textbf{S=-4}  &  $\Xi^{-} \Xi^{-}$  & $ \Xi^{-} \Xi^{0}  $  &  $\Xi^{0} \Xi^{0}$    & &  \\\hline
    \end{tabular}
\end{center}
\caption{Baryon-baryon channels for fixed strangeness $S$ and total charge $Q$.}
	\label{tab:particleplets}
\end{table}

\section{Interactions in Free Space}\label{sec:FreeSpace}
On the tree-level the octet interaction Lagrangians discussed in the previous section correspond to standard One Boson Exchange ($OBE$) amplitudes. In Tab. \ref{tab:OBE_Param} the OBE masses and cut-off momenta are displayed. They are kept fixed throughout the calculations, thus being treated as predetermined external input parameters. The full interaction amplitudes require the solution of the Bethe-Salpeter equation. As in \cite{Dejong:1998,Machleidt:1989}, we reduce the problem to an effective Lippmann-Schwinger equation by projecting out the time-like components by the Blankenbecler-Sugar method. However, different from the better known SU(2) nucleon-nucleon case, the full $SU(3)$ problem leads to a set of coupled equations describing interactions within the set of channels of fixed strangeness $S$ and total charge $Q$, respectively, where the latter corresponds to fix the third component of the total isospin, as discussed e.g. in \cite{Lenske:2015,Nijmegen,Dhar:2015} and shown in Tab. \ref{tab:particleplets}. In momentum representation and after performing a partial wave decomposition we have to solve the coupled integral equation
\be\label{eq:Tmat}
T_{ab}(q,q')=\tilde{V}_{ab}(q,q')+\sum_c{\frac{1}{2\pi^2}\int{dkk^2V_{ac}(q,k)Q_F(k,s,k_{F_c})G_c(k,s)T_{cb}(k,q') } }
\ee
where the indices $a,b,c$ account for all relevant quantum numbers in the corresponding flavour and partial wave channels, thus including a summation over the set of coupled baryon channels. $G_c$ denotes Green's function in the intermediate channel $c$, to be evaluated at the on-shell total center-of-mass energy, fixed by the Mandelstam variable $s$. In nuclear matter, the Pauli-projector
\be\label{eq:Q_F}
Q_F(k,s,k_{F_c})=\Theta(k^2_1-k^2_{F_1})\Theta(k^2_2-k^2_{F_2})
\ee
accounts by the step-functions for the Pauli-principle by blocking the occupied states inside the Fermi spheres $k_{F_c}=\{k_{F_1},k_{F_2}\}$ of particle 1 and particle 2 in channel $c$. In free space $Q_F$ reduces to the unity operator. Additional in-medium modifications are introduced by self-energies, mainly modifying the propagators. In-medium interactions will be discussed in more detail in later sections.

The Born-term contributes only at energies above the kinematical thresholds $s_{thr}=s_a,s_b$ where for a reaction $1+2\to 3+4$ the threshold energies are $s_a=(M_1+M_2)^2$ and $s_b=(M_3+M_4)^2$, respectively. Thus, we find
\be
\tilde{V}^{BB'}(q,q')=V(q,q')\Theta(s-s_{a})\Theta(s-s_{b})
\ee
thus accounting explicitly for $SU(3)$ mass symmetry breaking. In Tab. \ref{tab:Baryons} the baryon masses are given and the threshold energies are listed in Tab. \ref{tab:threshold}. For a more detailed understanding of the structure of the coupled equations and in order to clarify the invariant $\{8\}\otimes\{8\}$ multiplets from the given SU(3)-multiplet, it is useful to arrange the octet baryons into afore mentioned $BB'$ multiplets of conserved total strangeness and charge, Tab. \ref{tab:particleplets}.

By a properly chosen quadrature method Eq.~(\ref{eq:Tmat}) is conveniently transformed into a numerically easy to solve linear system, see e.g. \cite{Machleidt:1989}. The $BB'$ scattering matrix elements are determined numerically in by expressing the T-matrix in terms of the K-matrix, $T=\left(1-iK  \right)^{-1}K$ which is standard technique in reaction theory \cite{Machleidt:1989}. The K-matrix obeys a system  of equations as in Eq.~(\ref{eq:Tmat}) but using only the Cauchy Principal Value part of the intermediate propagators. Thus, the numerical problem reduces to the solution of a linear system of equations with real coefficients. The bare coupling constants $\left\{g_{D,F,S}\right\}_{P,S,V}$ are determined from fits to data, as discussed below. At present, the scarce data base, however, puts severe constraints on the precision of the derived values.

\begin{table}
\centering
\begin{tabular}{|l|r|l|r|l|r|}
\hline
$BB'$ & $\sqrt{s_{thres}}$ [MeV]& $BB'$ & $\sqrt{s_{thres}}$ [MeV]& $BB'$ & $\sqrt{s_{thres}}$ [MeV] \\ \hline
nn                     & 1879.13 & $\Xi^{-}n$             & 2261.27 & $\Sigma^{+}\Sigma^{0}$ & 2381.92 \\ \hline
np                     & 1877.84 & $\Sigma^{-}\Lambda$    & 2313.12 & $\Xi^{-}\Sigma^{-}$    & 2519.15 \\ \hline
pp                     & 1876.54 & $\Sigma^{-}\Sigma^{0}$ & 2389.99 & $\Xi^{-} \Lambda$      & 2437.39 \\ \hline
$\Lambda$ p            & 2053.96 & $\Lambda \Lambda$      & 2231.37 & $\Xi^{0}\Sigma^{-}$    & 2512.29 \\ \hline
$\Lambda$ n            & 2055.25 & $\Xi^{0}n$             & 2254.42 & $\Xi^{-}\Sigma^{0}$    & 2514.26 \\ \hline
$\Sigma^{+}p$          & 2127.64 & $\Xi^{-}p$             & 2259.98 & $\Xi^{0} \Lambda$      & 2430.54 \\ \hline
$\Sigma^{+} n$         & 2128.94 & $\Sigma^{0} \Lambda$   & 2308.23 & $\Xi^{0}\Sigma^{0}$    & 2507.41 \\ \hline
$\Sigma^{0}p$          & 2130.82 & $\Sigma^{0}\Sigma^{0}$ & 2385.10 & $\Xi^{-}\Sigma^{+}$    & 2511.08 \\ \hline
$\Sigma^{0}n$          & 2132.12 & $\Sigma^{-}\Sigma^{+}$ & 2386.81 & $\Xi^{0}\Sigma^{+}$    & 2504.23 \\ \hline
$\Sigma^{-}p$          & 2135.71 & $\Xi^{0}p$             & 2253.13 & $\Xi^{-}\Xi^{-}$       & 2643.42 \\ \hline
$\Sigma^{-}n$          & 2137.00 & $\Sigma^{0}p$          & 2253.13 & $\Xi^{-}\Xi^{0}$       & 2636.57 \\ \hline
$\Sigma^{-}\Sigma^{-}$ & 2394.87 &$\Sigma^{+}\Lambda$     & 2308.23 & $\Xi^{0}\Xi^{0}$       & 2629.72 \\ \hline
\end{tabular}
\caption{Baryon-baryon channel threshold energies in the particle basis.}
	\label{tab:threshold}
\end{table}

For a two-body reaction, the tree-level interactions are Lorentz-scalars built of vertex functionals and meson propagators,
\beq\label{obe}
V^{BB'}_{\alpha}(q',q)=g_{B_1B_3\alpha} \rho_{B_1B_3\alpha}(q,q')) D_\alpha(Q)
 \rho^\alpha_{B_2B_4}(-q,-q'))g_{B_2B_4\alpha}
\eeq
where in- and out-states are denoted by $B=(B_1,B_2)$ and $B'=(B_3,B_4)$, respectively. We have introduced the density matrices
\be
\rho_{BB'\alpha}(k,k')=\bar{u}^{B}(k')\kappa_\alpha u^{B'}(k) \quad .
\ee
Dirac spinors are indicated by $u^B(k)$. $\kappa_\alpha$ indicates the Dirac and flavour structure of the vertex. The meson propagators $D_\alpha$ depend on the 4-momentum transfer $Q$ which is either in the $t=(k'-k)^2$ or the $u=(k'+k)^2$ channel, respectively. In the two-body center-of-momentum frame, the 4-momenta in the in- and outgoing channels $(B_1,$ $B_2)$ and $(B_3,$
$B_4)$, respectively, are $k,k'$ with $k=(\omega,\bm{k})$. On the mass-shell, $k^2=q^2=m^2_B$ and the space-like components $\bm{q}$ are given by the total Mandelstam-energy $s_{12}=(p_1+p_2)^2$,
\be
q^2=\frac{1}{4s_{12}}\left( s_{12}-(M_1+M_2)^2 \right)\left( s_{12}-(M_1-M_2)^2 \right)\\
\ee
and $\bm{q}'$ is fixed accordingly. The meson propagators include vertex form factors of dipole type. The meson masses and the form factor cut-off momenta are found in Tab. \ref{tab:OBE_Param}. There, also the singlet-octet mixing angles are given which are taken into account in our calculations. Although in principle the values of the mixing angles are fixed by the mass relation discussed below in section \ref{sec:Octet_Interactions}, a breadth of values is in use for free-space calculations, trying to optimize the results. For the present purpose, we follow this strategy to some extent by using mixing angles in close orientation on the values of the Nijmegen group, extracting from the various interaction versions the values given in Tab. \ref{tab:OBE_Param} as appropriate averages, held fixed in the calculations.

\begin{figure}[tbh]
\begin{center}
\includegraphics[width=12cm]{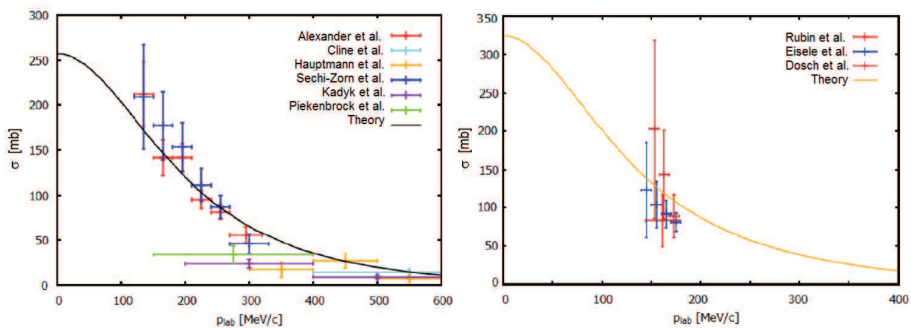}
\caption{Total cross section for elastic scattering of $\Lambda$ (left) and $\Sigma$ (right) hyperons on a hydrogen target as a function of the momentum of the incoming hyperon in the laboratory frame (see also \protect\cite{Dhar:2015}). Our theoretical results are compared to world-data set of experimental date from \protect\cite{Piekenbrock:1964,Cline:1967,Alex:1968,Sech:1968,Kadyk:1971,Hauptmann:1977} for $\Lambda+p$ and
\protect\cite{Dosch:1966,Rubin:1967,Eisele:1971} for $\Sigma+p$ cross sections, respectively.
}
\label{fig:LambdaSigma_xs}
\end{center}
\end{figure}

The tree-level matrix elements $V^{BB'}=\sum_\alpha{V_{BB'\alpha}}$ are entering as kernels into the Bethe-Salpeter equation. The latter is solved in three-dimensional reduction, resulting in an equivalent effective Lippmann-Schwinger equation for the K- or R-matrix, see e.g. \cite{Dejong:1998},
\bea\label{eq:BbS}
&&R^{BB'}(q',q)=V^{BB'}(q',q)\nonumber \\ &+&
\sum_{B''}{P\int{\frac{d^3k}{(2\pi)^3}V^{BB''}(q',k)G_{B''}(k,s)Q_F(k,s,k_{F_{B''}})R^{B''B'}(k,q)}}\quad .
\eea
which accounts for the ladder-resummation of two-body interactions in free space and infinite nuclear matter. The integration is treated as a Cauchy-Principal Value integral, as indicated. Since in free space $Q_F$ reduces to the unity operator. The reduction to an equivalent Lippmann-Schwinger equation in three dimensions is performed in the Blankenbecler-Sugar scheme. The two-particle propagator becomes in Blankenbecler-Sugar representation,
\be\label{eq:GBbS}
G_{bBS}(k,s)=4\sqrt{\frac{M_1M_2}{E_1(k)E_2(k)}}\frac{1}{(E_1(k)+E_2(k))^2-s} \quad ,
\ee
to be used for positive energy states only. Hence, in free space the operator $R$ reduces to the usual $K$-matrix from which phase shifts and scattering amplitudes are calculated.

\subsection{Hyperon-Nucleon Cross Sections}\label{ssec:Xsections}

Irrespective of the fact that there are several optimized parameter sets exist, we have decided by consistency reasons to determine a new set of coupling constants by a fit to the existing about 35 data points for total cross sections of $\Lambda+p$ and $\Sigma+p$ elastic scattering. Our strategy was to maintain the $SU(3)$ relations. Thus, the nine fundamental $SU(3)$ coupling constant $\{g_D,g_F,g_S\}$ for the pseudo-scalar, the vector, and the scalar meson nonets were used as fit parameters by a standard $\chi^2$-procedure \cite{MINUIT}. With $\chi^2=13.6$ the quality of the fit is far away from those obtained in the $NN$-case, where OBE-approaches like the Bonn-model typically lead to $\chi^2\simeq 1$. However, in view of the large uncertainties of the data base such a result had to be expected. During the fit, slight variations of the form factor cut-off masses by $\Delta \sim 100\cdots 200$~MeV had to be allowed for optimization. The resulting parameter set, including singlet-octet mixing, is displayed in Tab.\ref{tab:gDFS}. The values of the coupling constants in the various $BB'$-meson channels are easily found by means of the relation given in Eq.(\ref{eq:goct}).

\begin{table}[!h]
\begin{center}
\begin{tabular}{|l|c|c|c|c|}
  \hline
  Channel & $g_D$ & $g_F$ & $g_S$ & $\theta$ $[deg.]$ \\ \hline
  pseudo-scalar &  3.22016 & 2.14678 & 0.1913 & -23.0 \\  \hline
  vector & 0 & 1.48492 & 1.6037 & 35.26 \\  \hline
  scalar & 0.18809 & 1.15541& 1.8546 & 37.50\\
  \hline
\end{tabular}
\end{center}
\caption{The fundamental SU(3) coupling constants and the octet-singlet mixing angles for the three meson-nonets.}\label{tab:gDFS}
\end{table}

Our result are illustrated by comparison to the existing total elastic cross section data. In Fig. \ref{fig:LambdaSigma_xs} results for $\Lambda+p$ and $\Sigma+p$ cross sections are displayed. The few data points are seen to be described satisfactorily, but clearly, the data do not constraint sufficiently well the model parameters. This a well known finding in other approaches, too, see e.g. \cite{Nijmegen,Juelich,Epelbaum:2008}. The extrapolated threshold behaviour is typical for s-wave scattering. In the shown energy range actually only a few higher partial waves contribute to the cross section. At higher energies, just outside the shown window, the $\Lambda+p$ cross section shows the typical kink-behaviour at $p_{lab}\simeq 630$~MeV/c where the $\Sigma+p$ channels opens. A more extended sample of cross sections will be given elsewhere \cite{Dhar:2016}.

\section{Baryon-Baryon Interactions in Infinite Nuclear Matter}\label{sec:BB_Infinite}

The extension to in-medium scattering is readily achieved on the basis of the T-matrix, Eq.~(\ref{eq:Tmat}). In a medium, an important role is played by the Pauli-projector $Q_F$, defined in Eq.(\ref{eq:Q_F}). The 3-momenta $k_{i}$, entering into $Q_F$, are defined in the nuclear matter frame. They depend on the Mandelstam energy $s$ through the conserved total momentum, $\bm{P}^2=s-(E_1+E_2)^2$, since
\be
\bm{k}_{1,2}=\pm \bm{k}+x_{1,2}\bm{P}
\ee
with the Lorentz-invariant transformation coefficients
\be
x_{1,2}=\frac{s\mp (m^2_2-m^2_1)}{2\sqrt{s}}
\ee
where the upper sign correlates with particle 1 and $x_1+x_2=1$. Thus, we find in slightly different notation
\be
Q_F(k,P,u,k_F)=\Theta(k^2+x^2_1P^2+2x_1kPu-k^2_{F_1})\Theta(k^2+x^2_1P^2-2x_1kPu-k^2_{F_2})
\ee
where $u=cos(\beta)$ describes the angle between the vectors $\mathbf{k}$ and $\mathbf{P}$. In practice,  we use the angle-averaged projector
\be
\bar{Q}_F(k,P,k_F)=\frac{1}{2}\int^{+1}_{-1}{du Q_F(k,P,u,k_F)}
\ee
In nuclear matter, vector and scalar self-energies, $\Sigma^\mu=(\Sigma^0,\bm{k}E_i/M_i)^T$ and $\Sigma_s$, respectively, should be added, thus modifying the energies $E_i$ in the propagators Eq.~(\ref{eq:GBbS}). Thus, there are various sources of in-medium modifications. Among them, the Pauli-projector, however, provides typically the largest single contribution, as one easily sees from the Dyson-equation
\be
G^*_B(q,k_F)=G_B(q)-G_B(q)\Sigma_B(q,k_F) G^*_B(q,k_F) \quad ,
\ee
relating the in-medium single-particle Green-function $G^*_B(q,k_F)$ to the free-space propagator $G_B(q)$, depending only on the 4-momentum $q$. Hence, self-energy effects are suppressed $G\Sigma\sim \Sigma_B/M_B \ll 1$. Similar conclusions will be found for self-energies contributions in the in-medium spinors.

\begin{figure}[tbh]
\begin{center}
\includegraphics[width=12cm]{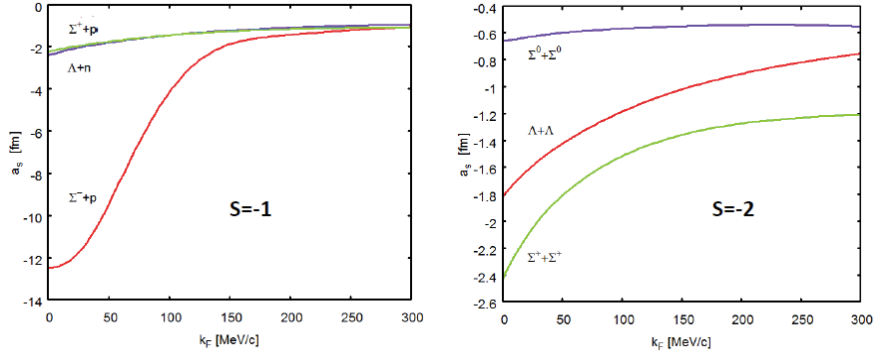}
\caption{In-medium Singlet-Even ($SE$) s-wave scattering lengths for the $S=-1$ hyperon-nucleon channels (left) and for the $S=-2$ double-hyperon channels (right) as a function of the Fermi-momentum $k_F\sim \rho^{1/3}$ (see also \protect\cite{Dhar:2015}).}
\label{fig:ScattLength}
\end{center}
\end{figure}

The low-energy parameters are an appropriate measure for interactions. For s-wave scattering at $q \to 0$ they are defined by
\be
\frac{1}{q}\tan{\delta_{s}(q,k_F)}\sim -a_{s}(k_F)\left(1+\frac{1}{2}q^2 a_{s}(k_F)r_{s}(k_F)+\cdots\right)
\ee
where $a_{s}$ and $r_{s}$ denote the scattering length and the effective range parameter, respectively. The scattering length is related to the volume integral of the R-matrix,  $a_{s}(k_F)=\frac{2\mu}{\hbar^2 4 \pi}R(0,k_F)_{|\ell=0}$, where $\mu$ is the reduced mass. Corresponding relations are found in all interaction channels. Thus, up to a normalization factor the scattering lengths correspond to in-medium effective coupling constants. In Fig. \ref{fig:ScattLength} the scattering lengths of the singlet-even ($SE$) channels are displayed for the channels with strangeness $S=-1,-2$.  With increasing density, the interaction strength decreases rapidly in all channels approaching a plateau at densities close to nuclear equilibrium. Overall, the scattering lengths are only about 10\% of the values found in the $NN$-system, for the latter see e.g. \cite{Machleidt:1989}, reflecting the much weaker $YY$ interaction. In fact, we encounter a known severe  problem: Using the $SU(3)$ parameters derived from $NY$-scattering for $NN$-scattering, the resulting $NN$ scattering lengths and cross sections are far off the experimental values.

With the T- and K-matrices available, we have access to the full $BB'$ scattering amplitude. For nuclear structure work we are interested in the first place on the G-matrix as introduced by Brueckner which is nothing but the in-medium K-matrix. While the natural representation of the K~-matrix is the \textit{singlet/triplet-even/odd} formalism, for nuclear matter calculations the explicit representation in terms of spin-isospin operators is more convenient. The two representation are related by an orthogonal transformation among the two operator sets. Thus, for each isospin doublet introduced in Eq.~(\ref{eq:doublets}) we obtain in non-relativistic reduction a structure familiar from the nucleon sector,
\bea
R^{BB'}(\xi,k_F)&=&
\sum_{S,I=0,1}{R^{BB'}_{SI}(\xi,k_F)\left(\bm{\sigma}_B\cdot\bm{\sigma}_{B'}\right)^S\left(\bm{\tau}_B\cdot\bm{\tau}_{B'}\right)^I  } \nonumber \\ &+&\sum_{I=0,1}{\left(R^{BB'}_{LI}(\xi,k_F)\mathbf{L\cdot S}+R^{BB'}_{TI}(\xi,k_F)\mathbf{S_{12}}\right)\left(\bm{\tau}_B\cdot\bm{\tau}_{B'}\right)^I }
\eea
and also spin-orbit and rank-2 tensor terms are displayed. The amplitudes $R_{SI}(\xi,k_F)$ are given either in momentum space ($\xi=\{\mathbf{k_1,k_2}\}$) or, if a static potential picture is an acceptable approximation, in coordinate space ($\xi=\{\mathbf{r_1,r_2}\}$). For a given $(BB')$-combination some of the $R_{SI}$ may vanish. Formally, also the interactions involving the $\Sigma$~-isotriplet can be cast into such a form, except that one (or both) Pauli-type isospin operators have to be replaced by the proper isospin-1 operator,  $\bm{\tau}_B\to \bm{T}_\Sigma$. In a single-hyperon spin-saturated nucleus the hyperon mean-fields are then given by
\be
U^{YA}=U^{YA}_{0}+U^{YA}_{L0}\bm{\ell_Y}\cdot\bm{\sigma_Y}+\left(U^{YA}_{1}+U^{YA}_{L1}\bm{\ell_Y}\cdot\bm{\sigma_Y}\right) \tau^0_Y
\ee
where the central isoscalar and isovector potentials $U^{YA}_{0,1}(\rho_{0,1})$ are determined by the isoscalar and isovector nuclear densities, $\rho_{0,1}=\rho_n\pm\rho_p$, respectively. Similar expressions hold also for the spin-orbit terms.

In infinite nuclear matter the hyperon-potentials have a particular simple structure. The static, momentum-independent part is determined in leading order by the s-wave scattering length. Taking as an example the $\Lambda$-case, we express the low-energy in-medium K-matrix in terms of the singlet and triplet spin-projectors $P_{1,3}$ and the $N\Lambda$ reduced mass $\mu_{N\Lambda}$
\bea
K_{\Lambda N}(q,k_{F})\simeq \frac{4\pi\hbar^2}{2\mu_{N\Lambda}}\left\{a^{SE}_{\Lambda N}(k_{F})P_{1}+a^{TE}_{\Lambda N}(k_{F})P_{3} \right\}
\eea
which leads to the nuclear matter potential
\be
U_\Lambda(\rho)=\frac{4\pi\hbar^2}{2\mu_{N\Lambda}}\left(\frac{1}{4}a^{SE}_{\Lambda N}(\rho)+\frac{3}{4}a^{TE}_{\Lambda N}(\rho) \right)\rho_N \quad ,
\ee
where $\rho$ is the total baryon number density. Corresponding potentials will result from $\Lambda Y$ interactions.

\section{Covariant DFT Approach to Hypernuclear Physics}
\label{sec:HyperDFT}

In the previous section, the attempt to study $YN$ and $YY$ interactions by a fit to the $YN$-scattering data was only partly successful in the sense that $NN$-scattering is not described properly by the derived set of $SU(3)$ coupling constants. In this section, we approach the problem from the reverse side: We start from well understood $NN$ quantities and use the $SU(3)$ relations to extrapolate into the strangeness sector. Since our primary interest is on hyperons in nuclear matter, we consider throughout in-medium interaction, based on our previous $NN$ DBHF results in asymmetric nuclear matter.

For obvious reasons, in nuclear structure research the meson fields giving rise to condensed classical field components are of primary importance. Hence, the isoscalar and isovector self-energies from scalar and vector mesons are dominating nuclear mean-field dynamics. In this section, devoted to nuclear ground state dynamics and single particle motion we therefore set the focus on the mean-field producing scalar and vector mesons. However, since we apply (Dirac-)Brueckner theory, the full set of mesons will contribute indirectly through the solution of the in-medium Lippmann-Schwinger equation. Self-energies will be evaluated formally on the Hartree-level but anti-symmetrization effects are included effectively by extracting density dependent coupling constants from the full DBHF/BHF self-~energies, as discussed in detail in our former work \cite{Keil:2002,Hofmann:2001,Lenske:2004,Dejong:1998}.

From the $1^-$ vector meson octet condensed isoscalar $\omega$ and isovector $\rho$ meson fields will evolve. In a system with a large fraction of hyperons also condensed scalar-octet $\kappa$ and vector-singlet $\phi$ mesons fields can appear. An apparent shortcoming of a pure SU(3) approach, however, is the absence of a clearly defined, natural scalar meson multiplet and, hence, the absence of a binding mean-field. As mentioned already in section \ref{sec:OctetInter}, we introduce a nonet of scalar mesons, treated as sharp mass states, but keeping in mind the inherent limitations.

Here, we use the DDRH Lagrangian including the baryon octet \cite{Keil:2002,Lenske:1995,Fuchs:1995,Hofmann:2001,Lenske:2004}.
In this case, the interaction Lagrangian is given by the isoscalar octet-mesons $\sigma$ and $\omega$ and their singlet $s\bar s$ counterpart $\sigma'$ and $\phi$, respectively. Isovector self-energies are described by the scalar $\delta/a_0(980)$ and the vector
$\rho$ mesons, respectively. Thus, we use the interaction Lagrangian, relevant for the mean-field sector,
\bea\label{eq:DDDR_Lagrangian}
\Lcal_{int} &=& \Psib_F \hat{\Gamma}_\sigma(\Psib_F, \Psi_F) \Psi_F \Phi_\sigma
- \Psib_F \hat{\Gamma}_\omega(\Psib_F, \Psi_F) \gamma_\mu \Psi_F V^\mu_\omega \nonumber \\
&&- \Psib_F \hat{\bm{\Gamma}}_\rho(\Psib_F, \Psi_F) \gamma_\mu \Psi_F
\vec{V}^\mu_\rho
+ \Psib_F \hat{\Gamma}_{\sigma'}(\Psib_F, \Psi_F) \Psi_F \Phi_{\sigma'} \nonumber\\
&&- \Psib_F \hat{\Gamma}_\phi(\Psib_F, \Psi_F) \gamma_\mu \Psi_F V^\mu_\phi
- e \Psib_F \hat{Q} \gamma_\mu \Psi_F V^\mu_\gamma,
\eea
which has to be complemented by the previously introduced baryon and meson kinetic Lagrangian densities $\Lcal_B$ and $\Lcal_M$.
The photon field is included and $\hat{Q}$ is the electric charge operator. The field-theoretical structure of the vertices as functionals of the matter field operators has been indicated explicitly. Of course, the full theory, as introduced before, includes the pseudo-scalar mesons as well. On the mean-field level, their contributions are contained in the vertex functionals which also account for anti-~symmetrization effects such that we finally end up with an easy to handle effective Hartree-theory. Formally, this is achieved by a Fierz-transformation, mapping the operator structure of the anti-~symmetrized u-channel interaction to an effective t-channel Hartree operator structure. In a last step, the u-channel propagators are averaged over the Fermi-sea. Thus, the u-channel is formally eliminated and mapped to the t-channel, extending our approach of Ref. \cite{Hofmann:1998} into the covariant regime. Obviously, the approach must be used only in a Hartree-scheme because otherwise uncontrollable double-counting effects will spoil the results.

\subsection{In-medium baryon-baryon vertices}
\label{ssec:BBvert}

In order to understand the subtleties of including baryon-baryon (BB)
correlations into a field theory of a higher flavour content we briefly
sketch the derivation of density dependent vertices $\Gamma$ from Dirac-Brueckner
theory. A Lagrangian of the type as defined above leads to a ladder kernel
$V^{BB'}(q',q)$ given in momentum representation by the superposition of
one boson exchange (OBE) potentials $V^{BB'\alpha}$(q',q).

\begin{figure}[tbh]
\begin{center}
\includegraphics[width=6cm]{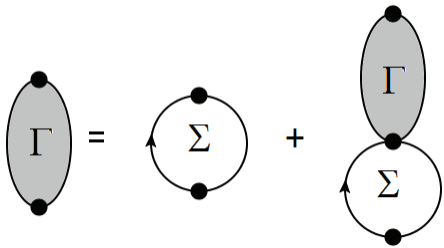}
\caption{Diagrammatic structure of the correlated vertex functionals $\Gamma$ in terms of the baryon self-energies $\Sigma$, normalized to the baryon mass (see text). Bare coupling constants $g^2$ are indicated by full circles.}
\label{fig:VertexDiag}
\end{center}
\end{figure}

In structure calculations the G-matrices $ R^{BB'}$ are required
in the nuclear matter rest frame rather than in the 2-body c.m. system.
In practice, the transformation is achieved by projection on the
standard set of scalar (S), vector (V), tensor (T), axial vector (A) and
pseudo scalar (P) Lorentz invariants \cite{HS:1983,HS:1987,HM:1987}. For our
purpose, however, a more convenient representation is obtained by
expanding the $R$~-matrices in terms of OBE-type propagators $D_a$ with masses $m_a$, as e.g. done in \cite{Hofmann:1998}.
Hence, we use the \emph{ansatz}
\be\label{eq:Separable}
R^{BB'}(q,q',k_F)=\sum_{a}{\Gamma^\dag_{a,\mu}(q_s,k_F)M^{B_1B_3}_a D_a(q,q')M^{B_2B_4}_a\Gamma^\mu_{a}(q_s,k_F)}
\ee
where $M^{B_iB_k}_a$ accounts for the spin-flavour structure of the vertices $a$. The momentum structure is covered by the propagators $D_a$. The dependence on the center-of-mass energy and the density of the background medium, expressed by $q_s$ and $k_F$, respectively, is contained in the vertex functionals $\hat{\Gamma_a}=\Gamma_a M_a$.
We define $\Gamma^2_a=\hat{\Gamma}^\dag_a\cdot \hat{\Gamma}_a$ and inserting this \emph{ansatz} into Eq.~(\ref{eq:BbS}),
we find the formal solution
\be\label{eq:Corr_Vertex}
\Gamma^2_a=D^{-1}_a\left[1- \int{dq'VGQ_F} \right]^{-1}V
\ee
where a symbolic notation has been used. By means of
\be
D^{-1}_a\to\frac{\delta}{\delta D_a}\quad ; \quad \frac{\delta V(q')}{\delta D_a(q)}=g^2_a\delta(q,q')
\ee
we obtain
\be
\Gamma^2_a=\frac{1}{(1-\int{VGQ_F})^2}g^2_a\left(1-\int{dq'VGQ_F(1-\delta(q,q'))}  \right)
\ee
and by neglecting the off-shell integral, we obtain in leading order
\be
\Gamma_a\simeq \frac{1}{1-\int{dq' VGQ_F}}g_a\sim\frac{1}{1-\frac{1}{M_B}\Sigma_B(k_F)}g_a
\ee
where $\Sigma_B\sim\int{dq VGQ_F}$ denotes the tree-level self-energy of the baryons $B$ on which the vertex operator is acting. Thus, the ladder series has been resummed with the result that energy and density dependent effects are mapped into correlated vertices. Moreover, the result provides access to systematic investigations of in-medium interactions by a (self-consistent) series expansion in terms of self-energies. The diagrammatic structure is depicted in Fig. \ref{fig:VertexDiag}.

The vertices $\hat{\Gamma}$ are carrying an implicit operator structure which becomes clear by considering that the matrix elements $M^{BB'}_a$ are in fact matrix elements of bilinears of fermion field operators, contracted with Dirac- and isospin/flavour-operators. This gives access to a field-theoretical interpretation, considering the correlated vertices as functionals of the baryon field operators $\hat{\Gamma}_a(\bar \Psi_B \Psi_B)$.

\begin{figure}[tbh]
\begin{center}
\includegraphics[width=10cm]{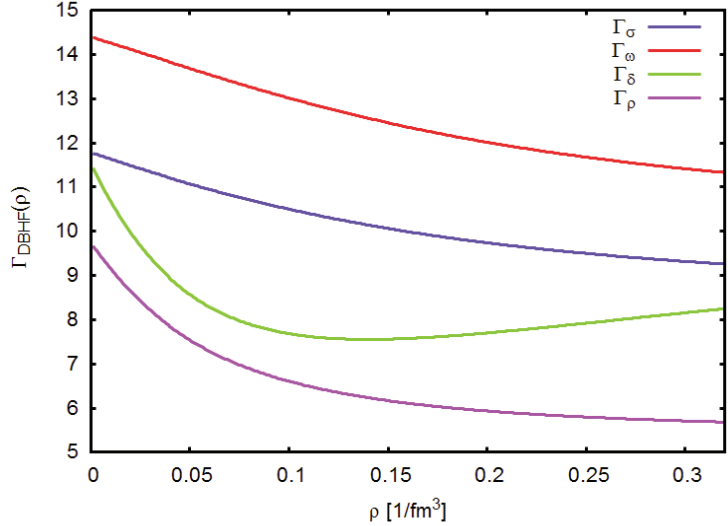}
\caption{In-medium $NN\alpha$ isoscalar ($\alpha=\sigma,\omega$) and isovector ($\alpha=\delta,\rho$) vertices determined from DBHF self-energies in asymmetric nuclear matter as explained in \protect\cite{Dejong:1998}.}
\label{fig:Gamma_DBHF}
\end{center}
\end{figure}

The full baryon self-energies have to be evaluated in terms of the G-matrix, i.e. the in-medium R-matrix. By means of Eq.~(\ref{eq:Separable}) we find
\be
\Sigma_{aB}(k,k_F)=tr_s tr_f\int{\frac{d^3k'}{(2\pi)^3}\Gamma^2_a(k',k_F)M^{B,B_f}_aD_a(k,k') n_{sB_f}(k',k_F)}
\ee
where traces over spin $s$ and flavours $f$ are indicated. The Fermi-momentum is denoted by $k_F$ and $n_{sB_f}(k)\sim \Theta(k_{F_B}-k)$ denotes the ground state occupation number of particles of given spin-flavour structure. Neglecting the (weak) momentum dependence of self-energies for the moment, we find the useful relation
\be\label{eq:SelfV}
\Gamma^2_{aB}(k_F)= \frac{1}{\rho_a}\Sigma_{aB}(k_F)+\mathcal{O}((\frac{k_F}{M})^2)
\ee
where the next-to-leading order term is indicated explicitly. The density is defined as
\be
\rho_a=tr_str_f\int{\frac{d^3k'}{(2\pi)^3}M^{B,B_f}_aD_a(k,k') n_{sf}(k',k_F)} \quad .
\ee
In practice, we encounter the vector and scalar densities with $M^{BB}_v=1$ and $M^{BB}_s=M_B/E_B(k)$, respectively. In infinite nuclear matter they are given by
\be
\rho_B=\frac{N_s}{6\pi^2}k^3_{FB}\quad ; \quad \rho_{sB}=\rho_B f_s(\frac{k_{F_B}}{M^*_B})
\ee
where $N_s=2$ is the spin multiplicity and
\be
f_s(z)=\frac{3}{2z^3}(z\sqrt{(1+z^2)}-\log{(z+\sqrt(1+z^2))}) \quad .
\ee
$M^*_B=M_B-\Gamma_{sB}(k_F)\Phi_s(k_F)$  indicates the relativistic effective baryon mass due to the action of the total scalar field $\Phi_s$, including isoscalar and isovector components. Thus, if the self-energies are known independently, Eq.~(\ref{eq:SelfV}) serves to determine the equivalent density dependent vertices. In Fig. \ref{fig:Gamma_DBHF} isoscalar and isovector vertices from purely nucleonic DBHF calculations are displayed. The general behavior is a reduction of the coupling strength with density. An exception is the isovector-scalar vertex: in asymmetric nuclear matter the scalar densities and fields obey a set of coupled non-linear equation leading to the peculiar behaviour seen in Fig.~\ref{fig:Gamma_DBHF}.

\begin{figure}[tbh]
\begin{center}
\includegraphics[width=8cm]{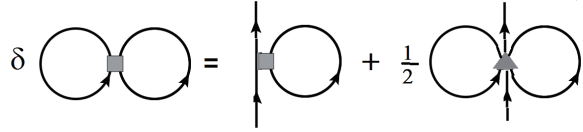}
\caption{Derivation of the self-energies by first variation of the energy density with respect to the density, leading to tadpole and rearrangement contributions, indicated by the first and second graph on the right hand side of the equation, respectively. The density dependent vertices are indicated by shaded squares, the derivative of the vertices by a triangle.}
\label{fig:SelfE}
\end{center}
\end{figure}

As a first application, the vertices are used in relativistic mean-field calculations, amounting to solve the stationary Dirac equation
\be
\left(\bm{\alpha}\cdot\bm{p}+\Sigma^0_B(\rho)+\gamma_0M^*_B(\rho)-e_{qB})  \right)\psi_{qB}=0
\ee
for a baryon $B$ in state $q$ with effective mass $M^*_B(\rho)=M_B-\Sigma^s_B(\rho)$, The vector and scalar self-energies are given by
\bea
\Sigma^0_B(\rho)&=&\Gamma^B_{\omega}(\rho)V^0_\omega(\rho)+\tau_B\Gamma^B_{\rho}(\rho)V^0_\rho(\rho)
+\Gamma^B_{\phi}(\rho)V^0_\phi(\rho)+\Sigma^{(r)}(\rho)\\
\Sigma^s_B(\rho)&=&\Gamma^B_{\sigma}(\rho)\Phi_\sigma(\rho)+\tau_B\Gamma^B_\delta(\rho)\Phi_\delta(\rho)+\Gamma^B_{\sigma'}(\rho)\Phi_{\sigma'}(\rho)
\eea
and for a finite nucleus the Coulomb potential $e_BV^0_\gamma(\rho)$ must be added to the vector self-energy,
where $e_B$ denotes the electric charge of particle $B$. The condensed fields $\varphi_\alpha\in \{ V^0_\omega,V^0_\rho,V^0_\phi,V^0_\gamma,\Phi_\sigma,\Phi_\delta,\Phi_{\sigma'} \}$ are obeying classical inhomogeneous Klein-Gordon field equations of the type
\be
\left(-\vec{\nabla}^2 + m^2_s  \right)\varphi_\alpha=\sum_B{\Gamma^B_\alpha(\rho)\rho^B_\alpha(\rho)} \quad .
\ee
In addition, rearrangement self-energies $\Sigma^{(r)}$ are contributing which are determined essentially by the density derivatives of the coupling constants \cite{Lenske:1995,Fuchs:1995,Lenske:2004}. They are accounting for vertex and propagator renormalization due to static polarization effects of the medium. In particular, the rearrangement self-energies are essential for ensuring basic thermodynamical relations \cite{Fuchs:1995}, most prominently found in the consistency relation for the thermodynamical and the mechanical pressure as expressed in the Hugenholtz-van Hove theorem \cite{Hugenholtz:1958}. The diagrammatic structure of the self-energies is illustrated in Fig. \ref{fig:SelfE}.

\begin{figure}[tbh]
\begin{center}
\includegraphics[width=12cm]{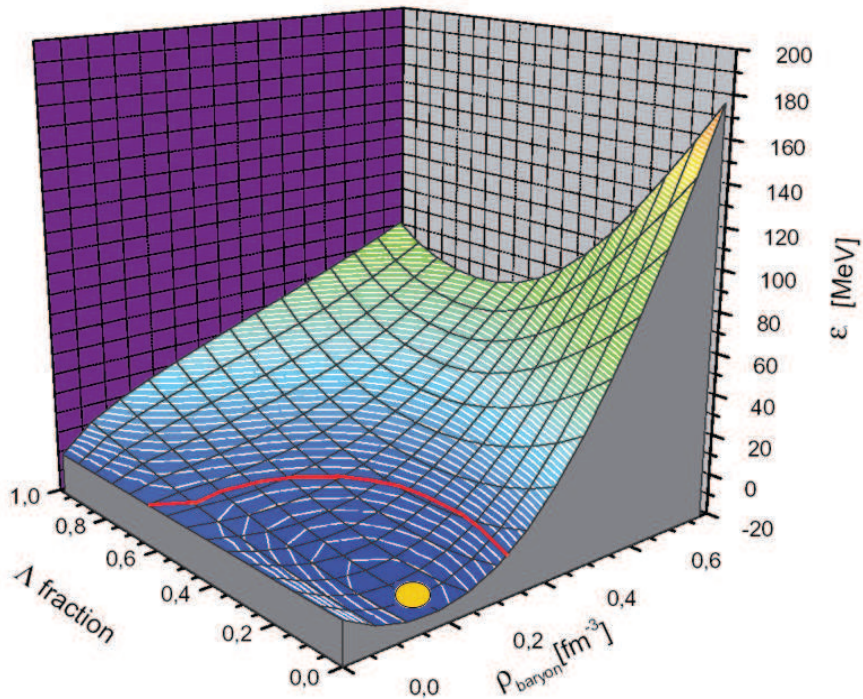}
\caption{Binding energy per baryon of ($n,p,\Lambda$) matter. The $\Lambda$ fraction is defined as $\xi_\Lambda=\rho_\Lambda/\rho$ and the background medium is chosen as symmetric ($p,n$) matter, $\xi_n=\xi_p$. The absolute minimum is marked by a filled circle. The line $\varepsilon=0$ is indicated by a red line.}
\label{fig:EoS-3D}
\end{center}
\end{figure}

In infinite matter, the field equations simplify to algebraic equations because the spatial derivatives vanish by symmetry. Then, the energy-momentum tensor
\bea
T^{\mu\nu} = \sum_i \frac{\p\Lcal}{\p(\p_{\mu}\phi_i)}\p^{\nu}\phi_i -
g^{\mu\nu}\Lcal
\\ \nonumber \phi_i \in \left\{\Psib,\Psi,V^\mu_{\omega},V^\mu_{\rho},V^\mu_{\gamma},V^\mu_{\phi},\Phi_{\sigma},
\Phi_{\delta},\Phi_{\sigma'}\right\}
\eea
can be evaluated in closed form. In infinite matter, the energy density and the pressure functionals are obtained from the energy-momentum tensor as
\bea
\epsilon &=& \langle T^{00} \rangle  =  \sum_{b=n,p}
    \frac{1}{4} \left[ 3E_{F_b}\rho_b + m_b^*\rho_b^s \right]
    \nonumber \\ & + &
    \frac{1}{2} \left[ m_{\sigma}^2\Phi_{\sigma}^2
                     + m_{\delta}^2\Phi_{\delta}^2
                     + m_{\sigma'}^2\Phi_{\sigma'}^2
                     + m_{\omega}^2{V^0_{\omega}}^2
                     + m_{\rho}^2{V^0_{\rho}}^2
                     + m_{\phi}^2{V^0_{\phi}}^2
                     \right] \nonumber \\
\label{eq:eNucMat} \\
p &=& \frac{1}{3}\sum_{i=1}^3\langle T^{ii} \rangle \nonumber \\
   &-& \sum_{B} \frac{1}{4} \left[ E_{F_B}\rho_B - M_B^*\rho_b^s
    \right] +
    \sum_{B} \rho_B\se^{0(r)}  \nonumber \\ & - &
    \frac{1}{2}\left[m_{\sigma}^2\Phi_{\sigma}^2
                   + m_{\delta}^2\Phi_{\delta}^2
                   + m_{\sigma'}^2\Phi_{\sigma'}^2
                   - m_{\omega}^2{V^0_{\omega}}^2
                   - m_{\rho}^2{V^0_{\rho}}^2
                   - m_{\phi}^2{V^0_{\phi}}^2
                   \right] \quad .
\label{eq:pNucMat}
\eea
Both are seen to be functionals of the number densities which in Lorentz-invariant form are given in terms of the baryon vector currents, $\rho^2_B=j_\mu j^\mu$. Infinite hypermatter, composed of protons, neutrons, and hyperons, is the natural extension of proton-neutron nuclear matter into the strangeness sector. We define the respective baryon fractions $\xi_B=\rho_B/\rho$, adding up to unity. As in the case of $(p,n)$-matter we consider the binding energy per particle
\be
\varepsilon(\rho)/\rho=\epsilon(\rho)/\rho-\sum_B{\xi_B M_B} \quad .
\ee
In Fig. \ref{fig:EoS-3D} results for $(p,n,\Lambda)$-matter are shown where $\Lambda$-hyperons are embedded into a background of symmetric $(p,n)$-matter. Hence, we fix $\xi_p=\xi_n$ and $\xi_\Lambda=1-2\xi_p$ such that at given total baryon number density $\rho_{p,n,\Lambda}=\xi_{p,n,\Lambda}\rho$. In Fig. \ref{fig:EoS-3D} the binding energy of that particular example of hypermatter is displayed.

The saturation properties of symmetric pure $(p,n)$-matter are very satisfactorily described: The saturation point is located within the experimentally allowed region at $\rho_{sat}=0.166fm^{-3}$ and $\varepsilon(\rho_{sat})=-15.95$~MeV with an incompressibility $K_\infty=268$~MeV which is at the upper end of the accepted range of values. Adding $\Lambda$ hyperons the binding energy first increases until a new minimum for 10\% $\Lambda$-content is reached at $\rho_{min}=0.21fm^{-3}$ with a binding energy of $\varepsilon(\rho_{min})=-18$~MeV. Increasing either $\xi_\Lambda$ and/or the density, the binding energy approaches eventually zero, as marked by the red line In Fig. \ref{fig:EoS-3D}. The minimum, in fact, is located in a rather wide valley, albeit with comparatively steep slopes, thus indicating the possibility of a large variety of bound single and even multiple-$\Lambda$ hypernuclei. Note, however, that the binding energy per nucleon considerably weakens at high densities as the $\Lambda$-fraction increases.

\subsection{Single-$\Lambda$ Hypernuclei}\label{ssec:FiniteLambda}

\begin{figure}[tbh]
\begin{center}
\includegraphics[width=12cm]{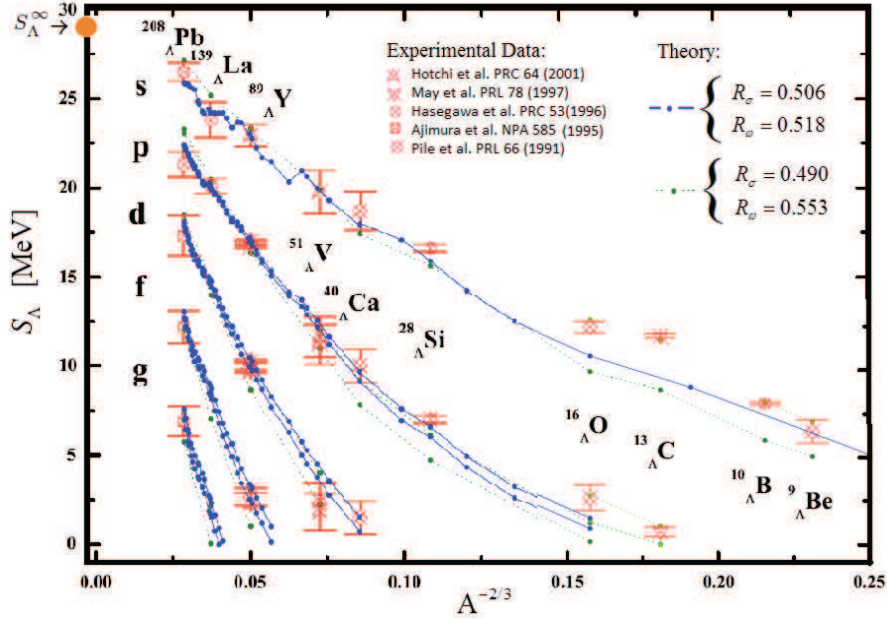}
\caption{Separation energies of the known $S=-1$ single $\Lambda$ hypernuclei as a function of the mass number $A$ to power $\gamma=-\frac{2}{3}$. Results of two sets of scaling parameter sets are compared to measured separation energies. For the $\ell >0$ levels the spin-orbit splitting is indicated. For $A\to \infty$ the limiting value $S^{\infty}_\Lambda \simeq 28$~MeV is asymptotically approached as indicated in the figure. Thus, we predict for the in-medium $\Lambda$-potential in ordinary nuclear matter a value of 28~MeV. The data are from refs. \cite{Hotchi:2001,May:1997,Hase:1996,Ajimura:1995,Pile:1991}.}
\label{fig:Lambda_Bnd}
\end{center}
\end{figure}

In the DDRH approach we have been following the widely used practice to relate $\Lambda$ coupling constants by a scaling factor to the corresponding nucleon coupling constants \cite{Keil:2002}. Hitherto unpublished results will be discussed in this section. The scaling hypothesis is based on the SU(6) quark model by assuming that the non-strange mean-fields couple only to the non-strange quark content of baryons. Na\"{\i}vely, this leads to the hypothesis that the $\Lambda$-vertices should be reduced globally by $1/3$, i.e. they are to be multiplied by a factor $R_\Lambda=\frac{2}{3}$. In practice, however, this is not confirmed by fits to $\Lambda$-hypernuclear spectra \cite{GL:1992}, partly because of global symmetry breaking on the mass scale and partly because of many-body effects \cite{Keil:2002}. Thus, in the DDRH approach we define
\be
\Gamma^\Lambda_{\sigma,\omega}(\rho)=R_{\sigma,\omega}\Gamma^N_{\sigma,\omega}(\rho)
\ee
and treat the scaling factors as free parameters. The investigations of infinite nuclear and neutron star matter, beta-stable and exotic nuclei \cite{Fuchs:1995,Hofmann:2001}, hypernuclei \cite{Keil:2002,Keil:2003}, and asymmetric nuclear matter at finite temperature \cite{Fedo:2014} have led to scaling factors ranging around $R_{\sigma\omega}\sim 0.5$ which, in fact, is close to the quark model hypothesis. In Fig. \ref{fig:Lambda_Bnd} the latest results for $\Lambda$ separation energies for the full set of known single-$\Lambda$ hypernuclei are shown. The calculations include also contributions from the relativistic tensor part. The KEK-data of Hotchi \textit{et al.} \cite{Hotchi:2001} for $^{41}_\Lambda V$ and $^{89}_\lambda Y$ have been especially important because of their good energy resolution and the observation of a large number of $\Lambda$ bound states.

\begin{table}
\begin{center}
  \begin{tabular}{|c|c|c|}
     \hline
     Level& $^{89}_{\Lambda}Y$ & $^{41}_{\Lambda}V$ \\ \hline
     $1s_{1/2}$ & -22.94 $\pm$ 0.64 MeV & -19.8 $\pm$ 1.4 MeV \\ \hline
     $1p_{3/2}$ & -17.02 $\pm$ 0.07 MeV & -11.8 $\pm$ 1.3 MeV\\ \hline
     $1p_{1/2}$ & -16.68 $\pm$ 0.07 MeV & -11.4 $\pm$ 1.3 MeV \\ \hline
     $1d_{5/2}$ & -10.26 $\pm$ 0.07 MeV & -2.7 $\pm$ 1.2 MeV \\ \hline
     $1d_{3/2}$ & -9.71 $\pm$ 0.07 MeV  &-1.9 $\pm$ 1.2 MeV\\ \hline
     $1f_{7/2}$ & -3.04 $\pm$ 0.11 MeV & $-$ \\ \hline
     $1f_{5/2}$ & -3.04 $\pm$ 0.11 MeV & $-$ \\ \hline
   \end{tabular}
   \caption{DDRH results for $\Lambda$ single particle energies.}\label{tab:Spectra}
\end{center}
\end{table}

A caveat for those nuclei is that the hyperon is attached to a high-spin core, $^{40}V(6^-)$ and $^{88}Y(4^-)$. Hence, the $\Lambda$ spectral distributions are additionally broadened by core-particle spin-spin interactions. A consistent description of the spectra could only be achieved by including those interactions into the analysis. A phenomenological approach was chosen by adding the core-particle spin-spin energy to the $\Lambda$ eigenenergies
\be
e_{(jJ_C)J\Lambda}=e^{RMF}_{j\Lambda}+E_{jJ_c}\langle (jJ_C)J|\bm{j}_\lambda\cdot\bm{J}_C|(jJ_C)J\rangle \quad .
\ee
giving rise to a multiplet of states. The multiplet-spreading is found to account for about half of the spectral line widths. Hence, if neglected, badly wrong conclusions  would be drawn on an extraordinary large spin-orbit splitting, too large by about a factor 2. Including the spin-spin effect leads to a spin-energy fully compatible with the values known from light nuclei. The analysis includes also the contributions from the relativistic tensor vertex \cite{Mares:1994}, modifying the effective $\Lambda$-spin-orbit potential to
\be
U^B_{so}=\frac{1}{r}\mathbf{r}\cdot\mathbf{\nabla}\left[\left(2\frac{M^*_B}{M_B}
\frac{f_{\Lambda\lambda\omega}}{g_{\Lambda\lambda\omega}}+1\right)\Sigma^\Lambda_\omega+\Sigma^\Lambda_\sigma  \right]
\ee
Here, the tensor strength $f$ appears as an additional parameter. For the $NN$ case, $f_{NN\omega}$ is known to be weak and usually it is set to zero. The small spin-orbit splittings observed in hypernuclei has led to speculations that the tensor part may be non-zero, partly cancelling the conventional spin-orbit potential, given by the difference of vector and scalar self-energies. This should happen for $f/g\sim -1$ as seen by considering that $U_{so}$ is a nuclear surface effect where $M^*\sim M$ and also the self-energies are about the same. The KEK-spectra are described the best for vanishing $\Lambda$ tensor coupling, $f_{\Lambda\lambda\omega}/g_{\Lambda\lambda\omega}=0$, thus agreeing with the $NN$-case. Our results for the $\Lambda$ single particle spectra in the two nuclei are found in Tab. \ref{tab:Spectra}.

The averaged spin-orbit splitting is about $223$~keV and $283$~keV and the spin-spin interaction amounts to $E_{jJ_c}=106$~keV and $E_{jJ_c}=61.3$~keV in Vanadium and Yttrium, respectively. Extrapolating the separation energies shown in Fig. \ref{fig:Lambda_Bnd} to (physical unaccessible) large mass number, the limiting value $S^{\infty}_\Lambda \simeq 28$~MeV is asymptotically approached for $A\to \infty$ which we identify with the separation energy of a single $\Lambda$~-hyperon in infinite nuclear matter. Thus, we predict for the in-medium $\Lambda$-potential in ordinary nuclear matter a value of 28~MeV.

\section{SU(3) Constraints on In-Medium Octet Coupling Constants}\label{sec:Octet_Interactions}

The $SU(3)$ relations among the coupling constants of the octet baryons and the $0^-,0^+,1^-$ meson nonets are conventionally used as constraints at the tree-level baryon interactions. In this section we take a different point of view. First of all, the mixing of singlet and octet mesons, to be discussed below, is considered. An interesting observation, closely connected to the mixing, is that in each interaction channel the three fundamental $SU(3)$ constants $g_D,g_F,g_S$ are already fixed by the $NN$ vertices with the isoscalar and isovector octet mesons and the isoscalar singlet meson, under the provision that the octet-singlet mixing angles are known. As shown below, the mixing angles depend only on meson masses. Since the Brueckner-approach retains the meson masses, the relations fixing the mixing angles are conserved by the solutions of the Bethe-Salpeter or Lippmann-Schwinger equation, respectively. Moreover, $SU(3)$ symmetry in general will be conserved, as far as interactions are concerned. The only substantial source of symmetry breaking is due to the use of physical masses from which one might expect $SU(3)$ violating effects of the order of 10\%. Although in low-energy baryon interactions the exchange particles are far off their mass shell, the on-shell mixing relations will persist because the $BB'$ T-matrices are symmetry conserving.

\begin{figure}[tbh]
\begin{center}
\includegraphics[width=10cm]{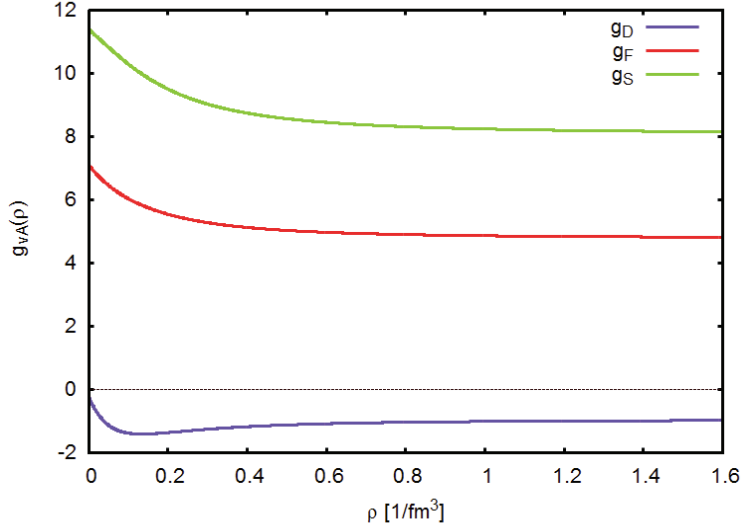}
\caption{In-medium fundamental $SU(3)$ vector vertices $g_{vA}$ (as indicated) versus the baryon density $\rho$. }
\label{fig:gDFS_vector}
\end{center}
\end{figure}
\subsection{Octet-Singlet Mixing}\label{ssec:OctSinMix}

In the quark model, the isoscalar mesons have flavour wave functions
\bea
|f_1\ra&=&\frac{1}{\sqrt{3}}\left( |u\bar u\ra + |d\bar d\ra + |s\bar s\ra \right)\\
|f_8\ra&=&\frac{1}{\sqrt{6}}\left( |u\bar u \ra + |d\bar d\ra -2|s\bar s\ra \right)\\
\eea
Since they are degenerate in their spin-flavour quantum numbers the physical states $f_8\to f$ and $f_1\to f'$ will be superpositions of the bare states. Taking this into account, the physical mesons are written as
\bea
|f_m \ra&=&\cos{\theta_m}|f_0\ra+\sin{\theta_m}|f_8\ra \\
|f'_m\ra&=&\sin{\theta_m}|f_0\ra-\cos{\theta_m}|f_8\ra
\eea
where $f_m\in \{\eta,\omega,\sigma \}$ and $f'_m\in \{\eta',\phi,\sigma_s \}$. \emph{Ideal mixing} is defined if $f_m$ does not contain a $s\bar s$ component and $f'_m$ is given as a pure $s\bar s$ configuration, requiring $\theta_{ideal}=\pi/2-\arcsin(\sqrt(2/3))\simeq 35.3$~°. \emph{Physical mixing} within a meson nonet, however, is reflected by mass relations of the Gell-Mann Okubo-type. The widely used linear mass relation \cite{PDG:2012}
\be
\tan{\theta_m}=\frac{4m_K-m_{a}-3m_{f'}}{2\sqrt{2}(m_{a}-m_K)}
\ee
leads to ($\eta,\eta'$) mixing with $\theta_P=-24.6$~° and ($\omega,\phi$)-mixing with $\theta_V=+36.0$~°, respectively. For the scalar nonet, the situation is less well understood, mainly because of the unclear structure of those mesons. The $0^{++}$ multiplets are typically strongly mixed with two- and multi-meson configurations or corresponding multi-$q\bar q$-configurations, leading to broad spectral distributions. For the purpose of low-energy baryon-baryon and nuclear structure physics we follow the successful strategy and identify the lowest scalar resonances as the relevant degrees of freedom. Thus, we choose the scalar octet consisting of the isoscalar $\sigma=f_0(500)$, the isovector $a_0(980)$ and the isodoublet $\kappa=K^*_0(800)$ mesons, respectively. While these mesons are observed at least as broad resonances \cite{PDG:2012}, the isoscalar-singlet partner $\sigma'$ of the $\sigma$-meson is essentially unknown. From the $SU(3)$ mixing relation, however, one easily derives the instructive mass relation \cite{PDG:2012},
\be
(m_f+m_{f'})(4m_K-m_a)-3m_fm_{f'}-8m^2_K+8m_Km_a-3m^2_a=0 \quad ,
\ee
serving as a constraint among the physical masses. Solving this equation for the scalar singlet meson we find the mass $m_{f'}=m_{\sigma'}=936^{+406}_{-88}$~MeV. The large uncertainty range indicates the uncertainties of the choice of the $\sigma$- and the $\kappa=K^*_0$-masses, mentioned before.  We use the mean mass values $M_\sigma=475$~MeV and $m_\kappa=740$~MeV, leading the scalar-singlet mass $m_{\sigma'}=936$~MeV, which lies close to mass of the $f_0(980)$ state, in good compliance with general expectations \cite{PDG:2012}. Then, the corresponding scalar mixing angle is $\theta_S=-50.73$~°. In Tab. \ref{tab:Tab2} the mixing results are collected.

\begin{figure}[tbh]
\begin{center}
\includegraphics[width=10cm]{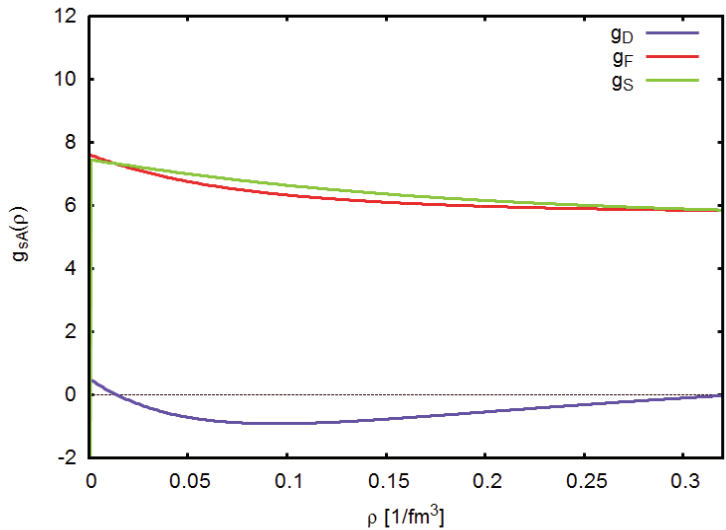}
\caption{In-medium fundamental $SU(3)$ scalar vertices $g_{sA}$ (as indicated) versus the baryon density $\rho$. }
\label{fig:gDFS_scalar}
\end{center}
\end{figure}

\begin{table}
\begin{center}
\begin{tabular}{|c|c|c|c|c|}
  \hline
  Channel & $f$ & $f'$ & $a$ & $\theta$ [deg.]\\ \hline
  Pseudo-scalar & $\eta$ & $\eta'$ & $\pi$ & -25.65 \\ \hline
  Vector & $\omega$ & $\phi$ & $\rho$ & +36.0 \\ \hline
  Scalar & $\sigma$ & $\sigma'$ & $a_0$ & -50.73 \\ \hline
\end{tabular}
\caption{Octet-Singlet meson mixing used to determine the in-medium vertices.}\label{tab:Tab2}
\end{center}
\end{table}

Meson-mixing affects directly the baryon interactions. The transformed $BB$-vector nonet coupling constants are displayed in Tab. \ref{tab:Couple_Mix} and corresponding relations hold for the pseudoscalar $\{\omega,\rho,K^*,\phi\} \to \{\eta,\pi,K,\eta'\}$ and the scalar nonets, $\{\omega,\rho,K^*,\phi\} \to \{\sigma,a_0,\kappa,\sigma'\}$.

\begin{table}[hbt]
\begin{center}
\begin{tabular}[b]{|r|l|}
\hline
Vertex & Coupling constant \\
\hline
$NN\omega$ &
$g_{NN\omega}=g_S \cos(\theta_v) +\frac{1}{\sqrt{6}} \left(3 g_F -g_D\right) \sin(\theta_v)$ \\ \hline
$NN\phi$ &
$g_{NN\phi}=g_S \sin(\theta_v) -\frac{1}{\sqrt{6}} \left(3 g_F -g_D\right) \cos(\theta_v)$ \\ \hline
$NN\rho$ & $g_{NN\rho}=\sqrt{2}(g_F+g_D)$\\ \hline
$\Lambda\Lambda\omega$ &
$g_{\Lambda\Lambda\omega}=g_S \cos(\theta_v) -\sqrt{\frac{2}{3}} g_D \sin(\theta_v)$\\ \hline
$\Lambda\Lambda\phi$ &
$g_{\Lambda\Lambda\phi}=g_S \sin(\theta_v) +\sqrt{\frac{2}{3}} g_D \cos(\theta_v)$\\ \hline
$\Sigma\Sigma\omega$ &
$g_{\Sigma\Sigma\omega}=g_S \cos(\theta_v) +\sqrt{\frac{2}{3}} g_D \sin(\theta_v)$\\ \hline
$\Sigma\Sigma\phi$ &
$g_{\Sigma\Sigma\phi}= g_S \sin(\theta_v) -\sqrt{\frac{2}{3}} g_D \cos(\theta_v)$\\ \hline
$\Sigma\Sigma\rho$ &
$g_{\Sigma\Sigma\rho}= \sqrt{2}g_F$\\ \hline
$\Lambda\Sigma\rho$ & $g^\rho_{\Lambda\Sigma}=\sqrt{\frac{2}{3}}g_D$\\ \hline
$\Xi\Xi\omega$ &
$g_{\Xi\Xi\omega}=g_S \cos(\theta_v) -\frac{1}{\sqrt{6}} \left(3 g_F +g_D\right) \sin(\theta_v)$ \\ \hline
$\Xi\Xi\phi$ &
$g_{\Xi\Xi\phi}=g_S \sin(\theta_v) +\frac{1}{\sqrt{6}} \left(3 g_F +g_D\right) \cos(\theta_v)$ \\ \hline
$\Xi\Xi\rho$ &
$g_{\Xi\Xi\rho}=\sqrt{2}(g_F-g_D)$\\
\hline
\end{tabular}
\caption{$SU(3)$ relations for the $\omega,\rho$ and $\phi$ baryon coupling constants, relevant for the mean-field sector of the theory.}\label{tab:Couple_Mix}
\end{center}
\end{table}

\subsection{SU(3) In-medium vertices}

Since we are primarily interested in mean-field dynamics we consider in this section interactions in the vector and the scalar channels only. From DBHF theory we have available in-medium isoscalar and isovector $NN$-vector and $NN$-scalar vertices as density dependent functionals $\Gamma_\alpha(\rho)$, see Fig. \ref{fig:Gamma_DBHF}. Thus, the $SU(3)$ relation, Tab. \ref{tab:Couple_Mix}, lead to the set of equations for the vector sector
\bea\label{eq:SU3_vector}
\Gamma_\omega(\rho)&=&g_{vS} \cos(\theta_v) +\frac{1}{\sqrt{6}} \left( 3 g_{vF} -g_{vD} \right )\sin(\theta_v) \nonumber \\
\Gamma_\phi(\rho)  &=&g_{vS} \sin(\theta_v) -\frac{1}{\sqrt{6}} \left( 3 g_{vF} -g_{vD} \right )\cos(\theta_v)  \nonumber \\
\Gamma_\rho(\rho)  &=&\sqrt{2}(g_{vD}+g_{vF})
\eea
and accordingly for the scalar sector,
\bea\label{eq:SU3_scalar}
\Gamma_{\sigma }(\rho)  &=&g_{sS} \cos(\theta_s) +\frac{1}{\sqrt{6}} \left( 3 g_{sF} -g_{sD} \right )\sin(\theta_s)  \nonumber \\
\Gamma_{\sigma'}(\rho)  &=&g_{sS} \sin(\theta_s) -\frac{1}{\sqrt{6}} \left( 3 g_{sF} -g_{sD} \right )\cos(\theta_s) \nonumber  \\
\Gamma_{\delta}(\rho)   &=&\sqrt{2}(g_{sD}+g_{sF})
\eea
For the present discussion we assume that the $NNf'$-singlet vertices $\Gamma_{\phi,\sigma'}$ vanish as it would be the case in the quark-model under ideal mixing conditions. Obviously, this constraint is easily relaxed and generalized scenarios with non-vanishing $NNf'$ coupling can be investigated.
\begin{figure}[tbh]
\begin{center}
\includegraphics[width=12cm]{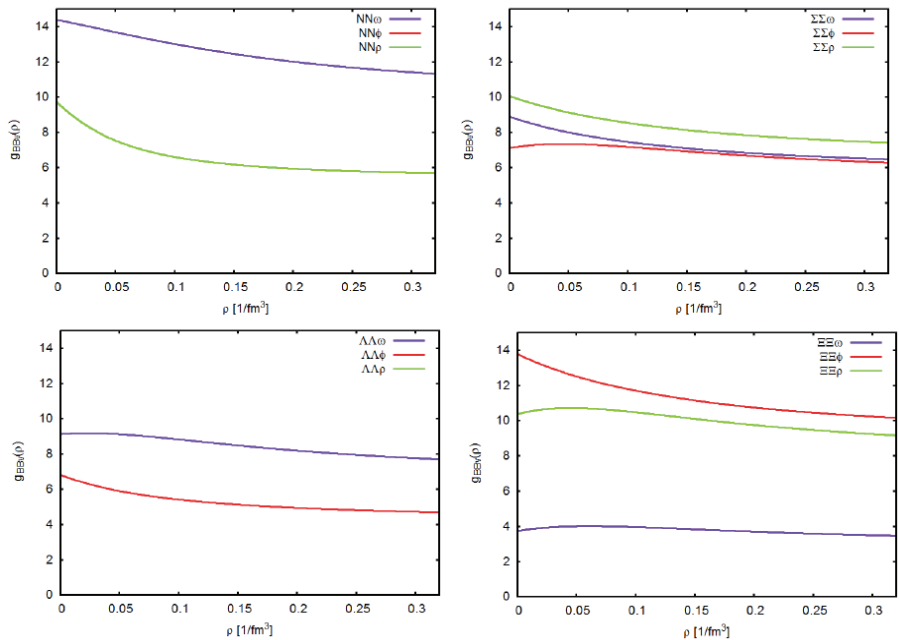}
\caption{In-medium $SU(3)$ vector vertices. The $NN$ and $\Lambda\Lambda$ vertices are found in the left column, the $\Sigma\Sigma$ and $\Xi\Xi$ vertices are displayed in the right column. Note that the $NN\phi$ and the $\Lambda\Lambda \rho$ coupling constants vanish identically.}
\label{fig:SU3_BB_vector}
\end{center}
\end{figure}

The resulting $SU(3)$ vertices are displayed in Fig. \ref{fig:gDFS_vector} and Fig. \ref{fig:gDFS_scalar} for the vector and the scalar nonets, respectively. In both the vector and the scalar channel, $g_D$ is found to be negative and with a modulus smaller than $g_F,g_S$ by a factor 5 to 10 which is in surprisingly good agreement with the general conclusion that $g_D$ should be small. However, here we derive this result from an input of coupling constants which describe perfectly well infinite nuclear matter and nuclear properties. The vertex functionals depend only weakly on the density with variations on the 10\% level over the shown density range.
\begin{figure}[tbh]
\begin{center}
\includegraphics[width=12cm]{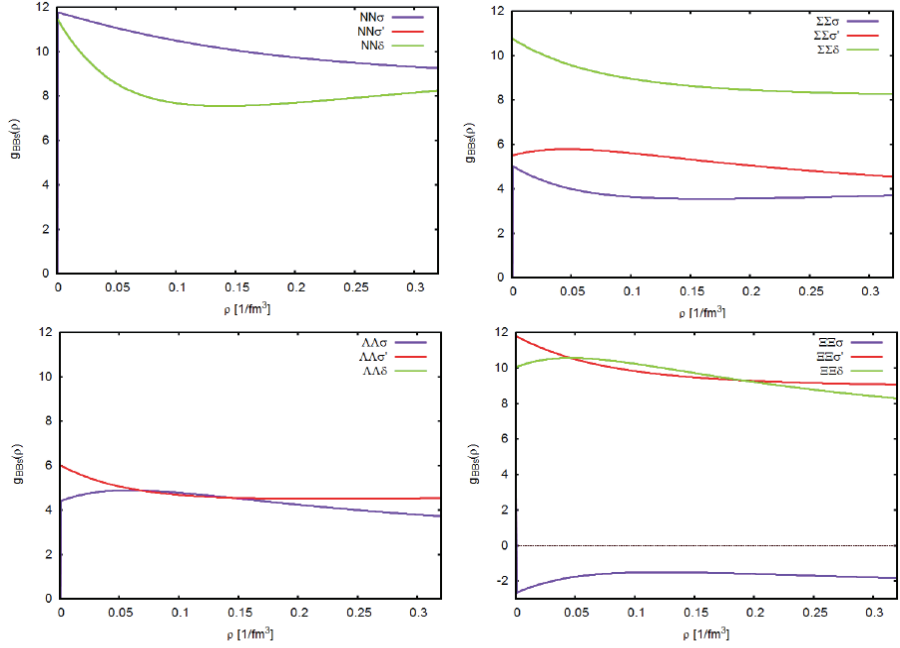}
\caption{In-medium $SU(3)$ scalar vertices. The $NN$ and $\Lambda\Lambda$ vertices are found in the left column, the $\Sigma\Sigma$ and $\Xi\Xi$ vertices are displayed in the right column. Note that the $NN\sigma '$ and the $\Lambda\Lambda \delta$ coupling constants vanish identically.}
\label{fig:SU3_BB_scalar}
\end{center}
\end{figure}
The resulting $BB$ vector and scalar vertices are shown in Fig. \ref{fig:SU3_BB_vector} and Fig. \ref{fig:SU3_BB_scalar}. The scaling hypothesis works quite well for the $\Lambda$-hyperon where up to saturation density indeed an almost constant value of about $R_{\omega,\sigma}\sim 0.5\cdot 0.6$ is found, also surprisingly close to the quark model estimate. Roughly the same situation is found for the isoscalar-octet vertices in the $\Sigma$- and $\Xi$~-channels: The isoscalar-octet $\Sigma$-vertex scaling factors agree to a good approximation with the values found for the $\Lambda$. The isoscalar-octet $\Xi$ scaling factors are ranging close to $0.25\cdots 0.3$ which is surprisingly close the quark-model expectation of $\frac{1}{3}$. However, the $\Xi\Xi\sigma$ vertex involves also a change of sign which would never obtained by the scaling hypothesis. The isovector interactions do not follow the na\"{\i}ve  quark-model scaling hypothesis. There, one finds scaling constants of the order of unity. In hypermatter with more than a single hyperon, sizable condensed isoscalar-singlet fields will evolve to which the $\Lambda,\Sigma,\Xi$-baryons will couple. The $\Xi$-interactions, for example, are dominated by the isoscalar-singlet and the isovector-octet channels which might shed new light on the dynamics of $S=-2$ hypernuclei.

\begin{figure}[tbh]
\begin{center}
\includegraphics[width=8cm]{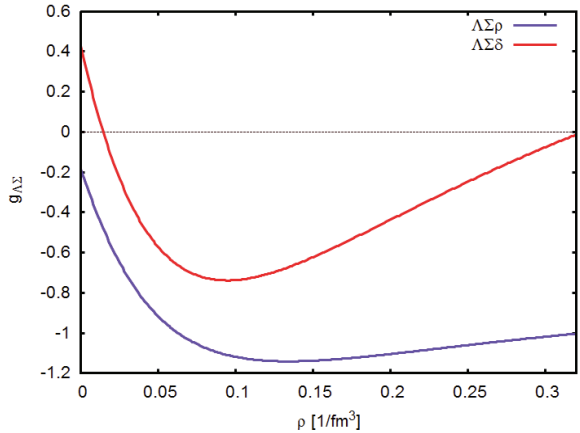}
\caption{In-medium $SU(3)$ $\Lambda$~-$\Sigma^0$ interaction vertices resulting from the isovector-vector and isovector-scalar interactions, respectively.}
\label{fig:LambdaSigma_mix}
\end{center}
\end{figure}

In Fig. \ref{fig:LambdaSigma_mix} we show the vertex functionals leading to the mixing of $\Lambda$ and $\Sigma^0$ hyperons in asymmetric hypermatter, as discussed in section \ref{ssec:SymBreak}. The resulting non-diagonal mixing self-energy will be obtained by the difference of the vector and the scalar partial contribution, obtained by multiplying the vertices of Fig. \ref{fig:LambdaSigma_mix} by the isosvector-vector and the isovector-scalar density, respectively.  
\section{Summary and Outlook}\label{sec:SumLook}
Strangeness physics is a field of particular interest for our understanding of baryon dynamics in the very general context of low-energy flavor physics. The well known $SU(3)$ scheme connects the various interaction vertices of octet baryons and meson multiplets and also here, we have exploited those relations serving to reduce the number of free parameters. A new set of interaction parameters for free space baryon-baryon interactions was derived and used in in-medium calculations. The density dependence of $YN$ and $YY$ interactions was investigated by considering the evolution of low-energy scattering parameters with increasing density of the background medium. The general trend of the scattering lengths reflects a rapid decrease of the effective interaction strengths at higher density. In detail, however, the density dependence is different for each of the various strangeness channels. This behaviour is driven to a large extent by the mass differences, thus being a consequence of $SU(3)$ symmetry breaking on the mass scale.  An open problem for the future is the consistent treatment of three- and many-body forces, urgently needed for neutron star physics, see e.g. \cite{Yama:2014}.

For systematic studies of nuclear properties, a density functional approach was presented, following closely and extending our previously introduced DDRH theory. Density dependent nucleon-meson vertices from DBHF theory are being used in a covariant approach. Results for infinite hyper matter and finite $\Lambda$-hypernuclei were discussed. In hypermatter the minimum of the binding energy per particle was found to be shifted by adding $\Lambda$ hyperons  to larger density and stronger binding. Using the same interaction in our covariant energy density functional, a good description of the known set of hypernuclear $\Lambda$-separation energies was obtained.

In a next step, we have exploited the $SU(3)$ relations in a completely different context. Starting from well established $NN$-DBHF results the $SU(3)$ relations were used to derive in-medium vector and scalar fundamental coupling constant $\{g_D,g_F,g_S \}$ and the corresponding vertices in the various $BB$ channels. For the first time, an approach was presented allowing to investigate the density dependence of the full set of $SU(3)$ coupling constants. We emphasize again that these vertex functionals are based on microscopically derived $NN$-vertices, describing nuclear properties very well. Since the input of DBHF-vertices are directly obtained from free-space $NN$-interactions, the resulting $BB$-vertices are of \emph{ab initio} nature. Applications of these interactions in nuclear matter and hypernuclear calculations are in preparation. An interesting case is the mean-field induced mixing briefly discussed in section \ref{ssec:SymBreak} which seems to have been overlooked largely.

An interesting result is that the quark-model scaling hypothesis is roughly recovered from the $BB$-vertices, calculated independently without imposing explicitly quark-model constraints. However, since the underlying $NN$-interactions were determined from data, they contain necessarily already implicitly the full information on $SU(3)$-relations, if they are of any importance for $BB$-interactions. Because the Dirac-Brueckner-equations are symmetry conserving, the DBHF-quantities still carry that information and confirm surprisingly well the seemingly \emph{ad hoc} assumptions of the scaling hypothesis.

Last but not least, the $BB$ interactions discussed here are also of interest for reaction studies. First of all, they may be used to describe bound states of hyperons, produced in any kind of reaction, e.g. the ones discussed elsewhere in this volume \cite{Gaitanos:2016}. Secondly, in-medium cross sections can be calculated to be tested in transport calculations.

\subsection*{Acknowledgements:}
\noindent Contributions by C. Keil are gratefully acknowledged. Supported by DFG, contract Le439/9, BMBF, contract 05P12RGFTE, GSI Darmstadt, and Helmholtz International Center for FAIR.

\end{document}